\documentclass[aps,pra,superscriptaddress,amsmath,amsfonts,amssymb,floatfix,nofootinbib]{revtex4}
\usepackage{enumerate}
\usepackage{mathtools}
\usepackage{amsmath}
\usepackage{graphicx}
\usepackage{epsfig}
\usepackage{sidecap}
\usepackage{hyperref,stackrel}
\usepackage[normalem]{ulem}
\usepackage{color}
\usepackage{enumerate}
\usepackage{bm}
\usepackage[utf8]{inputenc}
\usepackage{amsmath}
\usepackage{amsfonts}
\usepackage{amssymb}
\usepackage{graphicx}
\usepackage{enumitem}
\usepackage{physics}
\usepackage{tikz}
\usepackage{amsthm}
\newtheorem{theorem}{Theorem}
\newtheorem{lemma}{Lemma}
\newtheorem{proposition}{Proposition}
\newtheorem{definition}{Definition}
\theoremstyle{remark}
\newtheorem{remark}{Remark}
\newcommand{\rev}{}
\newcommand{\revii}{}
\newcommand{\Fsum}{\mathcal{F}_{\rm sum}}
\newcommand{\Fp}{\mathcal{F}_{\rm price}}
\newcommand{\Shi}{S_{\rm hi}}
\DeclareMathOperator{\rnd}{round}

\begin{document}

\title{{On the generic structures of the protocols for quantum auction and quantum
summation and their relation}}
\author{Kabir Sandhu } \thanks{kabirsandhu111@gmail.com}
\affiliation{Indian Institute of Technology, Ropar, India}
\author{Sandeep Mishra} \thanks{sandeep.mtec@gmail.com}
\affiliation{Jaypee Institute of Information Technology, A-10, Sector-62, Noida, UP-201309, India}
\author{Anirban Pathak} \thanks{anirban.pathak@gmail.com}
\affiliation{Jaypee Institute of Information Technology, A-10, Sector-62, Noida, UP-201309, India}

\begin{abstract}
Secure multi-party computation (SMC) addresses the problem of jointly computing global functions of private inputs while revealing minimal information about individual data. Two prominent examples of SMC tasks are sealed-bid auction and secure multi-party summation. Existing schemes for quantum auction and quantum summation have largely been developed independently, motivated by distinct applications and employing different computational primitives. {\rev In this work, mutual reductions between the two tasks are established for single-item sealed-bid auctions with first-price payment. In one direction, an auction used as a black box computes a secure sum: each party bids a private random key drawn from its input, so the clearing price depends on the inputs only through their sum $S$. The reduction is perfectly private against colluding semi-honest parties and uses $O(S^2\log(1/\delta))$ auction rounds or $O(S\log(1/\delta))$ under coherent access. It can be instantiated with an existing auctioneer-free quantum auction after a one-line change to its disjunction subroutine, which as published leaks occupancy counts. In the other direction, summation with randomized inputs yields threshold disjunctions, and a single search over composite keys locates the highest bid and a winner, revealing nothing else. Because each query consumes the bidders' encodings, a bidder can adapt its effective bid undetected, while a fixed query set restores sealed bidding. Combinatorial auctions and general payment rules lie outside this framework. On a superconducting processor, one instance of each direction is executed, recovering the winner for four-bidder instances with ties and the sum of private inputs from repeated auctions.}
\end{abstract}
\maketitle

\section{Introduction}

With the advent of quantum technologies, quantum computation and quantum communication have now emerged as two important areas of quantum science and technology \cite{preskill2018quantum,pirandola2020advances}. Traditionally, quantum computing drew the attention of the community because of the capability of quantum algorithms to perform certain computational tasks much faster than their classical counterparts \cite{Shor, Grover, QAlgo}. In contrast, the most important aspect of quantum communication is linked to its potential to provide unconditional security \cite{akpthk,portmann2022security}. However, these two verticals are not really disjoint. Interestingly, there are computational tasks where the security and privacy of the inputs provided by different participants play a crucial role. Such tasks are generally referred to as secure multi-party computation (SMC) tasks. In fact, SMC is a cryptographic paradigm in which a set of mutually distrusting parties, each holding a private input, jointly compute a public function of their inputs such that no party learns anything beyond what can be inferred from its own input and the prescribed output \cite{Crepeau2002,GMW1987,Yao1982,Lindell2017}. The SMC tasks have many practical applications, as the majority of the e-commerce-related activities, sealed-bid auction, voting, etc. can be viewed as SMC tasks. The fact that quantum technologies have the potential to outperform their classical counterparts led to the analysis and development of quantum versions of the SMC protocols. As a consequence, several schemes for quantum sealed-bid auction \cite{shi2021quantum,PathakAuction1,PathakAuction2}, quantum voting \cite{Vaccaro2007Voting,PathakVoting1}, quantum anonymous veto \cite{PathakVeto2,rahaman2015quantum,PathakVeto1}, quantum summation \cite{ji2019quantum,yang2018secure,shi2016secure}, quantum e-commerce \cite{PathakEcommerce}, quantum lottery \cite{mishra2023quantum}, etc. have been proposed in the recent past. In the present work, we wish to specifically focus on the schemes for quantum sealed-bid auction and secure multi-party summation. The rationale for focusing on these schemes is two-fold. Firstly, a sealed-bid auction is important in its own right, and if unconditionally secure schemes for sealed-bid auctions can be developed, then corruption can be reduced to a large extent, which will have an extremely positive impact on the growth of developing countries. Secondly, the schemes for the two tasks mentioned above seem to be related, and we are interested in exploring the relationship. Classical auction mechanisms, including first-price and Vickrey auctions, have motivated extensive work on secure and privacy-preserving protocols \cite{Goldreich2009SMC}. As mentioned above, recently, quantum auction schemes have been proposed to enhance privacy guarantees, reduce communication complexity, and exploit quantum resources such as superposition and interference \cite{PathakAuction1, PathakAuction2}.

A task that is closely related to sealed bid auction is secure multi-party summation, in which multiple parties jointly compute aggregate quantities such as totals or averages over private data \cite{shi2016secure,yang2018secure,ji2019quantum}. Secure summation lies at the core of multi-party computation and privacy-preserving data analysis, and {\rev quantum summation algorithms based on amplitude estimation, a notion distinct from secure multiparty summation protocols (see the Notation section), offer} a quadratic advantage over classical sampling methods \cite{brassard2002quantum,grinko2021iterative}. Despite the conceptual similarity between auctions and summation, these two tasks have largely been studied in isolation, using different abstractions and motivated by different application domains. Lately, a set of schemes for quantum secure multi-party summation have been proposed and analyzed \cite{dou2024quantum,li2024secure,zhang2024new,zhang2025novel}. All the schemes of summation and auction mentioned above are designed independently, and so far they are treated as two independent and separate facets of SMC tasks.

In this work, we show that {\rev for single-item sealed-bid auctions} this separation is unnecessary. In what follows, we demonstrate that auction protocols and summation protocols are {\rev mutually reducible} and both share a common operational structure. Our central observation is that {\rev for single-item sealed-bid auctions with scalar bids, threshold testing, maximum bid determination, and winner determination} can be reduced to repeated evaluations of summation oracles acting on appropriately defined indicator functions. {\rev Conversely, a sealed-bid auction used as an unmodified black box computes a sum. Each party submits the random bid $\lfloor B\,U_i^{1/x_i}\rfloor$ built from its private input $x_i$ and fresh private randomness $U_i$. The clearing price then satisfies $\Pr[b_{\max}<m]=(m/B)^{S}$ exactly, so the clearing prices of repeated rounds are a statistic of $S=\sum_i x_i$ alone. We make both directions explicit. Firstly, we show that repeated auctions disclosing only the clearing price compute $S$ with perfect privacy against colluding semi-honest parties {\revii (parties that follow the protocol but may pool their views in a coalition, as defined in Section~\ref{sec:setting} and in the threat model of Section~\ref{sec:security-analysis})} using $O(S^2\log(1/\delta))$ rounds{\revii, where $\delta$ is the failure probability of the protocol}, that this round count cannot be improved for independently sampled bids, that private randomness is necessary, that publishing the identity of the winner reveals every input, and that coherent access to the auction reduces the cost to $O(S\log(1/\delta))$ invocations through amplitude estimation. We instantiate the black box with an existing auctioneer-free quantum sealed-bid auction \cite{shi2022feasible} and show that its disjunction subroutine must be modified slightly to meet the privacy requirement.} {\rev Secondly, we show that one search over composite keys, whose queries are summations with randomized inputs, locates both the highest bid and a winner, reveals nothing else, and never compares bids with each other. Since quantum encodings are consumed by each query, the search forces bidders to re-encode after observing intermediate outcomes. We show that this lets a bidder condition its effective bid on the other bids without detection, give a sealed variant with a fixed query set, provide an error budget for approximate amplitude-estimation oracles, and compare the construction with quantum maximum finding.} These constructions are protocol-agnostic and apply across {\rev computational models, but only within this class of auctions}. {\rev Finally, we execute both directions on an IBM superconducting processor. The complete search returns the highest bid and the winner for five bid vectors, three of them with ties, and a control in which the register is dephased shows that the disjunction query depends on coherence. In the other direction, $470$ auctions built from the auctioneer-free protocol of Ref.~\cite{shi2022feasible} return the sum of the private inputs, and the leakage of its published label rule, together with its removal by the modification of Section \ref{sec:existing}, is measured on the same device.}

{\rev By identifying summation as the primitive underlying single-item sealed-bid auctions,} our results offer a modular perspective on secure multi-party quantum computation. This viewpoint enables hardware-independent protocol design and suggests new routes toward experimentally realizable, privacy-preserving quantum computation schemes. Before we elaborate on the findings of the present work, for the convenience of the readers, it will be apt to introduce the notation and the model used in this work.

{\rev
\subsection*{Scope and related work}

All results of this paper assume that a single item is sold, that each bidder holds one scalar bid in a linearly ordered set, and that the outcome consists of the highest bid, a winner, and the first-price payment. This assumption is what gives threshold indicators and the searches of Section \ref{summation_auction} their meaning, and it is what the exponential race of Section \ref{auction_summation} relies on. Several important settings fall outside, and they cannot be reached by extending the present framework. In a combinatorial auction with $M$ items a valuation is a set function on $2^M$ bundles, so there is no linear order on which to place thresholds, and winner determination is a weighted set packing problem, which is NP-hard \cite{rothkopf1998computationally,sandholm2002algorithm}, rather than a counting statistic. Summation aggregates independently across bidders and therefore discards the complementarities between items that make combinatorial auctions difficult, and the exposure problem \cite{milgrom2004putting,cramton2006combinatorial} cannot be expressed in its language. Payment rules other than first price, equilibrium bidding, revenue, efficiency, and multi-unit formats are likewise not analyzed, apart from the second-price remarks in Sections \ref{sec:reduction} and \ref{sec:search} and the sealed-bidding results of Section \ref{sec:strategy}.

The general case has been studied by Hogg, Harsha, and Chen \cite{hogg2007quantum}, who represent bids as superpositions and use a distributed adiabatic search to identify the winners. Their protocol covers first-price and second-price formats, single items and arbitrary bundles of items, and it lets bidders use entanglement to arrange correlations among their bids that other participants cannot observe. Measuring the final state gives the outcome and destroys the superposition, so non-winning bids are not revealed, and the authors analyze how a suitably designed adiabatic search limits the ability of bidders to game the auction. Laboratory experiments with human bidders \cite{chen2008experiments} and an analysis of a corrupt auctioneer with circuit implementations \cite{guha2008quantum} followed. Game-theoretic quantum auctions were also studied by Piotrowski and S\l{}adkowski \cite{piotrowski2003quantum,piotrowski2008quantum}. The present work covers a strictly smaller class of mechanisms, and bundle bidding and correlated bidding have no counterpart here. Its contribution is different in kind: explicit reductions in both directions between single-item first-price sealed-bid auctions and secure summation, stated as ideal functionalities with simulation-based privacy against colluding semi-honest parties, an optimal round complexity, an instantiation with an existing protocol, and the consequences of re-encoding for leakage and bidder strategy.

Quantum sealed-bid auctions in the cryptographic sense began with the protocol of Naseri \cite{naseri2009secure}, whose vulnerabilities \cite{qin2009cryptanalysis,yang2009improved} led to post-confirmation mechanisms \cite{zhao2010secure,liu2016multiparty} and to the circular key-agreement protocol of Ref.~\cite{PathakAuction1} and the semi-quantum protocol of Ref.~\cite{PathakAuction2}. Later protocols protect non-winning bids \cite{shi2019privacy,shi2021quantum}, anonymity \cite{shi2022anonymous}, or remove the auctioneer \cite{shi2022feasible}, and measurement-based and entanglement-phase variants have been proposed \cite{ye2023measurement,mihara2023quantum}. Classical privacy-preserving auctions range from distributed auction services \cite{franklin1996design} and auctioneer and issuer models \cite{naor1999privacy} to fully private bidder-resolved auctions \cite{brandt2006obtain}, deployed multiparty computation \cite{bogetoft2009secure}, and auctioneer-free protocols based on anonymous veto \cite{bag2020seal}. Protocols for secure multiparty quantum summation include those of Refs.~\cite{chen2010efficient,shi2016secure,yang2018secure}, with attacks and improvements reported in Refs.~\cite{gu2019improvement,zhang2019improvements}. Related quantum primitives include anonymous ranking \cite{huang2014quantum} and the quantum summation algorithms of numerical analysis \cite{heinrich2002quantum}, whose amplitude-estimation core underlies the Notation section below.
}

\subsection*{Notation and amplitude-estimation model}\label{sec:notation}

Let $\mathcal{H}_I$ denote the $N$-dimensional Hilbert space spanned by $\{|i\rangle : i=1,\dots,N\}$ and let
$\mathcal{H}_A$ denote a single ancilla qubit. We may represent the uniform superposition over indices as
\begin{equation}
    |u_N\rangle = \frac{1}{\sqrt{N}}\sum_{i=1}^N |i\rangle .
\end{equation}
A quantum summation primitive receives oracle access to a Boolean function $f:\{1,\dots,N\}\rightarrow\{0,1\}$ through a unitary
\begin{equation}
    U_f : |i\rangle|0\rangle \mapsto |i\rangle\big( \sqrt{1-f(i)}\,|0\rangle + \sqrt{f(i)}\,|1\rangle \big).
\end{equation}
Amplitude estimation applied to $U_f$ yields an estimate of
\begin{equation}
    \mu(f) = \frac{1}{N}\sum_{i=1}^N f(i),
\end{equation}
with $O(1/\epsilon)$ queries achieving additive error $\epsilon$, improving upon the classical $O(1/\epsilon^2)$ sample complexity \cite{brassard2002quantum}.

{\rev \emph{Two meanings of summation and the normalization.} The word summation is used in two senses. A secure multiparty summation protocol \cite{chen2010efficient,shi2016secure,yang2018secure} is a cryptographic functionality that outputs the exact sum of private integers or group elements, and it is what Sections \ref{sec:existing} and \ref{sec:disj} use. Quantum summation in the algorithmic sense \cite{heinrich2002quantum,brassard2002quantum} estimates $\mu(f)$ with the operator $U_f$, which requires $f(i)\in[0,1]$. The two are linked by $S=N\mu(f)$, so an additive error $\epsilon$ on $\mu$ is an error $N\epsilon$ on the sum. Section \ref{auction_summation} works with integer inputs $x_i\in\{0,\dots,X\}$, and real inputs in $[0,X]$, which include $f(i)\in[0,1]$, are covered by Remark \ref{rem:real}. Section \ref{summation_auction} never encodes bid values in amplitudes. It encodes only the Boolean indicators $f_t(i)=\mathbb{1}[\kappa_i\ge t]$, so no normalization by $B_{\max}$ enters the encoding. The normalization does enter the query cost. Deciding whether $S_t>0$ needs $O(\sqrt N)$ applications of $U_{f_t}$ (Proposition \ref{prop:budget}). Resolving the integer count $S_t$ exactly needs additive error below $1/(2N)$ on $\mu$, which costs $O(\sqrt{S_t(N-S_t)})$ and at most $O(N)$ applications \cite{brassard2002quantum}. Encoding the bids themselves as $f(i)=b_i/B_{\max}$ and estimating $\sum_ib_i$ to within $\pm1/2$ needs additive error below $1/(2NB_{\max})$, which costs $O(NB_{\max})$ applications.

\emph{Symbols.} $N$ is the number of parties or bidders. In Section \ref{auction_summation}, $x_i$, $X$, $S$, and $\Shi=NX$ are the summation inputs, their bound, their sum, and its bound, $B$ is the size of the bid space of the black-box auction, $M$ counts auction rounds, and $\Lambda$ is the dimension of an amplitude-estimation register. In Section \ref{summation_auction}, $b_i\in\{0,\dots,B_{\max}\}$ are bids, $\kappa_i$ are composite keys, and $L$ is the number of search levels. {\revii Here $B_{\max}$ is a given upper bound on the bids, which are the private inputs of the auction built in Section~\ref{summation_auction}, so it plays the role that $X$ plays for the inputs $x_i$ in Section~\ref{auction_summation}. It is distinct from $B$, the parameter of the black-box auction of Section~\ref{auction_summation} that also appears in the Introduction, which the parties choose and which must grow linearly with $\Shi$ (Theorem~\ref{thm:correct} takes $B\ge32e\,\Shi$). The two are related only when an auction built as in Section~\ref{summation_auction} serves as the black box of Section~\ref{auction_summation}, which requires $B_{\max}\ge B-1$. For a threshold $t$, $S_t=\sum_if_t(i)=N\mu(f_t)$ is the number of bidders whose key $\kappa_i$ reaches $t$, and Section~\ref{sec:disj} uses the same count with the bids $b_i$ in place of the keys. It is the sum $S$ above with each input $x_i$ replaced by the Boolean value $f_t(i)$, so that $X=1$ and $S_t\in\{0,\dots,N\}$.} Throughout, $\delta$ is a failure probability and $\epsilon$ a leakage or precision parameter, as stated where they appear.}

As the motivation for the present work, with emphasis on the relevance of SMC task is already provided and the notation to be used in this work is briefly summarized in this section, we may now move to the technical findings of this paper which is organized as follows.  {\rev In Section \ref{auction_summation}, we use a sealed-bid auction as a black box to compute a secure sum, prove correctness, privacy, and optimality of the round complexity, and give a coherent version based on amplitude estimation.} {\rev In Section \ref{summation_auction}, we locate the highest bid and a winner from summation queries, and analyze leakage, incentives, and approximation errors.} {\rev In Section \ref{sec:experiment}, we execute both directions on IBM quantum hardware.} This is followed by security analysis in Section \ref{sec:security-analysis}. The computational, communication and memory costs are analyzed in Section \ref{cost_analysis}. Finally, the work is concluded in Section \ref{sec:conclusion}. {\rev Proofs are collected in Appendix \ref{app:proofs}.}

\section{Summation from a sealed-bid auction} \label{auction_summation}

{\rev
In this section a sealed-bid auction is used as an unmodified black box, and a secure sum is recovered from its public output alone. The allocation rule and the payment rule of the auction are left untouched. The parties only decide what to bid.

\subsection{Setting}\label{sec:setting}

There are $N\ge 2$ parties $P_1,\dots,P_N$. Party $P_i$ holds a private integer $x_i\in\{0,1,\dots,X\}$, and the parties wish to compute
\begin{equation}
S=\sum_{i=1}^{N}x_i\in\{0,1,\dots,\Shi\},\qquad \Shi=NX .
\label{eq:sum}
\end{equation}
The ideal summation functionality $\Fsum$ receives $(x_1,\dots,x_N)$ and announces $S$.

The black box is a single-item first-price sealed-bid auction with bid space $\{0,1,\dots,B\}$. We describe it by an ideal functionality $\Fp$ that receives one sealed bid $b_i$ from every party, allocates the item to a highest bidder, charges the winner the clearing price $b_{\max}=\max_i b_i$, and announces $b_{\max}$. Ties are broken by a rule that depends only on the set of tied bidders and on independent randomness. We distinguish three disclosure variants. In $\Fp$ the identity of the winner is not disclosed to the parties, in $\Fp^{\rm priv}$ only the winner is notified, and in $\Fp^{\rm pub}$ the winner is announced publicly. Any quantum sealed-bid auction protocol that realizes one of these functionalities can be substituted for it, and the results below then hold for that protocol within the security guarantee that the protocol itself provides. Section \ref{sec:existing} identifies which existing quantum sealed-bid auctions qualify and instantiates the black box with the auctioneer-free protocol of Shi and Li \cite{shi2022feasible}.

Parties are semi-honest and may collude. A coalition $C\subsetneq\{1,\dots,N\}$ follows the protocol but pools its views. We write $H$ for the complementary set of parties and $S_C=\sum_{i\in C}x_i$, $S_H=S-S_C$. This adversary model is stronger than the no-collusion model of Section \ref{sec:security-analysis}.

\subsection{The reduction}\label{sec:reduction}

\emph{One auction round.} Every party $P_i$ draws fresh private randomness $U_i$ uniformly from $(0,1)$, forms the key
\begin{equation}
K_i=\begin{cases}U_i^{1/x_i}, & x_i\ge 1,\\ 0, & x_i=0,\end{cases}
\label{eq:key}
\end{equation}
and submits the bid $b_i=\lfloor B K_i\rfloor\in\{0,\dots,B-1\}$ to the auction. The public output of the round is the clearing price $p=b_{\max}$.

\begin{lemma}[Threshold law]\label{lem:threshold}
For every $m\in\{1,\dots,B\}$,
\begin{equation}
\Pr[p<m]=\left(\frac{m}{B}\right)^{S}.
\label{eq:threshold}
\end{equation}
Consequently, the distribution of the clearing price depends on the inputs only through $S$.
\end{lemma}
\begin{proof}
For $x_i\ge1$, $b_i<m$ holds if and only if $BK_i<m$, that is $U_i<(m/B)^{x_i}$, which has probability $(m/B)^{x_i}$. For $x_i=0$ the event has probability $1=(m/B)^{0}$. The bids are independent, so $\Pr[p<m]=\prod_i (m/B)^{x_i}$.
\end{proof}

The key \eqref{eq:key} is the exponential race. The variable $-\ln K_i$ is exponential with rate $x_i$, the maximum key corresponds to the minimum of independent exponentials, which is exponential with rate $S$, and party $P_i$ holds the largest key with probability $x_i/S$ \cite{efraimidis2006weighted}. Estimating sums from extrema of exponential variables is standard in distributed computing \cite{cohen1997size,mosk2008fast}. The new elements here are that the only operation applied to the keys is a sealed-bid auction, that discretization to a finite bid space leaves the threshold law \eqref{eq:threshold} exact, and the privacy, optimality, and coherent results below.

\begin{remark}[Payment rule of the black box]
The reduction needs an auction that publishes the highest bid. A second-price auction does not serve, because the second-highest bid does not satisfy Lemma \ref{lem:threshold}. Its threshold law is $\Pr[\text{at most one bid}\ge m]=\prod_iq_i\,\big[1+\sum_i(q_i^{-1}-1)\big]$ with $q_i=(m/B)^{x_i}$, which depends on the individual inputs and not only on $S$. For example, $x=(2,0)$ and $x=(1,1)$ give different laws.
\end{remark}

\emph{Estimator.} For a threshold $m$ let $c=-\ln(m/B)$, and let $\hat q$ be the fraction of $M$ independent rounds with $p<m$. By Lemma \ref{lem:threshold} each round contributes a Bernoulli variable with mean $q=e^{-cS}$. The estimate is
\begin{equation}
\hat S=\rnd\!\left(-\frac{\ln \hat q}{c}\right),
\label{eq:estimator}
\end{equation}
clipped to $[0,\Shi]$, with $\hat S=\Shi$ if $\hat q=0$.

\emph{Protocol $\Pi_{\rm race}$.} The public parameters are $\Shi$, $B$, and a failure probability $\delta$.
\begin{enumerate}[leftmargin=*]
\item \emph{Scale stage.} Let $J=\lceil\log_2\Shi\rceil+1$ and run $M_1=\lceil 14\ln(4(J+1)/\delta)\rceil$ rounds. For $j=0,\dots,J$ set $m_j=\lfloor Be^{-2^{-j}}\rfloor$ and let $\hat q_j$ be the fraction of these rounds with $p<m_j$. Let $j^\ast$ be the smallest $j$ with $\hat q_j\ge 0.335$ ($j^\ast=J$ if there is none) and set $\tilde S=\min(2^{j^\ast+1},\Shi)$.
\item \emph{Resolution stage.} Run $M_2=\lceil 52\,\tilde S^{2}\ln(4/\delta)\rceil$ fresh rounds, set $m=\lfloor Be^{-1/\tilde S}\rfloor$, and output $\hat S$ from Eq.~\eqref{eq:estimator}.
\end{enumerate}
The bids never depend on the thresholds. The thresholds enter only the classical post-processing of public prices.

\begin{theorem}[Correctness and round complexity]\label{thm:correct}
Let $B\ge 32e\,\Shi$. For every input with sum $S$, $\Pi_{\rm race}$ outputs $S$ with probability at least $1-\delta$. With probability at least $1-\delta/2$ the total number of auction rounds satisfies
\begin{equation}
M_1+M_2\le 14\ln\frac{4(J+1)}{\delta}+3328\,\max(S,1)^2\ln\frac{4}{\delta}+2,
\label{eq:rounds}
\end{equation}
and it never exceeds $M_1+\lceil 52\,\Shi^2\ln(4/\delta)\rceil$. The non-adaptive variant that omits the scale stage and sets $\tilde S=\Shi$ requires only $B\ge 8e\,\Shi$ and uses $\lceil 52\,\Shi^2\ln(2/\delta)\rceil$ rounds.
\end{theorem}

The proof is given in Appendix \ref{app:thm1}. The constants in Eq.~\eqref{eq:rounds} come from Hoeffding's inequality \cite{hoeffding1963} and are loose. Exact binomial computation shows that about $26\,S^2$ resolution rounds suffice for $\delta=0.05$ [Fig.~\ref{fig:reduction}(b)], and the success probability depends on $M$ and $S$ only through $M/S^2$ [Fig.~\ref{fig:reduction}(a)].

\begin{remark}[Size of the bid space]
The bid space must grow linearly with $\Shi$. If $S\gg B$, the clearing price equals $B-1$ except with probability $(1-1/B)^S\le e^{-S/B}$, and the number of rounds needed to resolve $S$ grows exponentially in $S/B$. With $B=\Theta(\Shi)$ each bid occupies $\lceil\log_2 B\rceil=\log_2\Shi+O(1)$ qubits or bits.
\end{remark}

\begin{remark}[Real-valued inputs]\label{rem:real}
Lemma \ref{lem:threshold} holds verbatim for real $x_i\ge0$. Replacing the rounding step by a multiplicative tolerance, the argument of Appendix \ref{app:thm1} gives a $(1\pm\epsilon)$ estimate of $S$ with $O(\epsilon^{-2}\log(1/\delta))$ resolution rounds once a constant-factor estimate of $S$ is available. This covers the normalized model $f(i)\in[0,1]$ of the Notation section.
\end{remark}

\subsection{Privacy}\label{sec:privacy}

The privacy argument uses a single property of the black box.

\begin{definition}[Max-bounded leakage]\label{def:maxleak}
An auction protocol on the bid space $\{0,\dots,B\}$ has max-bounded leakage if, for every coalition $C\subsetneq\{1,\dots,N\}$, the view of $C$ in one invocation, which includes the clearing price, is distributed as $L_C\big(b_C,\,\max_{i\in H}b_i,\,R\big)$ for a fixed map $L_C$ and randomness $R$ independent of all bids.
\end{definition}

The ideal functionality $\Fp$ has max-bounded leakage with $L_C=\max(\max_{i\in C}b_i,\,h)$, where $h=\max_{i\in H}b_i$. Protocols that locate the maximum through private disjunctions also qualify, because a coalition holding the winning bid may additionally learn leading bits of $h$ (Section \ref{sec:existing}).

\begin{theorem}[Perfect privacy]\label{thm:privacy}
Let the black box have max-bounded leakage, for example $\Fp$. For every coalition $C\subsetneq\{1,\dots,N\}$ there is a simulator that, given only $(x_C,S)$, outputs a sample distributed exactly as the joint view of $C$ in $\Pi_{\rm race}$, namely its inputs, its randomness, its views of all auction invocations, and the output. Hence $\Pi_{\rm race}$ reveals nothing beyond $S$ and is a perfectly private protocol for $\Fsum$ against colluding semi-honest parties, with correctness error $\delta$.
\end{theorem}
\begin{proof}
In each round the honest keys are independent of the coalition's randomness. Applying Lemma \ref{lem:threshold} to the honest parties alone, $h=\max_{i\in H}b_i$ satisfies $\Pr[h<m]=(m/B)^{S_H}$, and $S_H=S-S_C$ is computable from $(x_C,S)$. The simulator draws the coalition's randomness honestly, draws $h$ from this law and $R$ from its distribution, and outputs $L_C(b_C,h,R)$. Rounds are independent, the number of rounds is a function of earlier prices, and the output is a function of the prices, so the simulated and real joint distributions coincide.
\end{proof}

If max-bounded leakage holds only up to computational indistinguishability, as for SEAL \cite{bag2020seal} under the decisional Diffie-Hellman assumption, the same simulator yields a computationally indistinguishable view.

\begin{proposition}[Private winner notification]\label{prop:priv}
Let the black box be $\Fp^{\rm priv}$ and let $M_{\max}=M_1+\lceil52\,\Shi^2\ln(4/\delta)\rceil$. Extend the simulator of Theorem \ref{thm:privacy} by reporting a coalition win exactly when $\max_{i\in C}b_i>h$, with ties inside the coalition broken by the public rule. The simulated view is within total variation distance $M_{\max}S/B$ of the real view. Choosing $B\ge M_{\max}\Shi/\epsilon$ bounds the leakage by $\epsilon$ at the cost of $O(\log\Shi+\log(1/\epsilon)+\log\log(1/\delta))$ additional bid qubits.
\end{proposition}
\begin{proof}
The two views can differ only in rounds with $\max_{i\in C}b_i=h$ and $S_H\ge1$. If $S_H=0$, all honest bids are $0$ and the tie structure is known to the simulator. For $S_H\ge 1$ the probability of such a round is at most $\max_j\Pr[h=j]=\max_j\big[((j+1)/B)^{S_H}-(j/B)^{S_H}\big]\le S_H/B$ by the mean value theorem. A union bound over at most $M_{\max}$ rounds completes the proof.
\end{proof}

\begin{proposition}[Public winner identity reveals every input]\label{prop:pub}
Let the black box be $\Fp^{\rm pub}$ with uniform tie breaking and $B\ge\max(32e\,\Shi,\,4N\Shi^2)$, and let $n_i$ be the number of resolution rounds won by $P_i$. With probability at least $1-\delta-2N(\delta/4)^{6.5}$ the public transcript yields $S$ together with every individual input through $x_i=\rnd(S\,n_i/M_2)$.
\end{proposition}
\begin{proof}
With continuous keys $P_i$ wins with probability $x_i/S$ \cite{efraimidis2006weighted}. Discretization changes the winner only if another bidder shares the top bin, which for each bidder $k$ requires $b_k=\max_{j\ne k}b_j$, an event of probability at most $S/B$ by the argument of Proposition \ref{prop:priv}. The winning probability $w_i$ therefore satisfies $|w_i-x_i/S|\le NS/B\le 1/(4S)$. On the event $\tilde S\ge S$, which has probability at least $1-\delta/2$, we have $M_2\ge 52S^2\ln(4/\delta)$, and Hoeffding's inequality gives $|n_i/M_2-w_i|<1/(4S)$ except with probability $2\exp(-M_2/8S^2)\le 2(\delta/4)^{6.5}$. Then $|S n_i/M_2-x_i|<1/2$. A union bound over the $N$ parties and over the correctness event of Theorem \ref{thm:correct} completes the proof.
\end{proof}

Theorem \ref{thm:privacy} and Propositions \ref{prop:priv} and \ref{prop:pub} imply that whether a given quantum sealed-bid auction protocol can serve as the black box is decided by what it discloses, and not by its internal mechanism. Protocols that announce the winner cannot be used directly. In a Monte Carlo test with $x=(3,0,5,2)$, $\Shi=10$, $B=1600$, and $\delta=0.05$, the public winner counts recovered every individual input exactly in $300$ out of $300$ runs.

\subsection{An existing quantum auction as the black box}\label{sec:existing}

Theorem \ref{thm:privacy} and Propositions \ref{prop:priv} and \ref{prop:pub} turn the choice of black box into a checkable condition. The protocol must release nothing beyond a function of the coalition's own bids and the highest honest bid. In particular it must not reveal individual bids, counts of bidders in a bid interval, or the identity of the winner. The quantum sealed-bid auctions closest to this work fail this condition as specified. In the protocol of Sharma et al.~\cite{PathakAuction1} every party decodes every bid at the final measurement. In the semi-quantum protocol of Asagodu et al.~\cite{PathakAuction2} the auctioneer decodes all bids before announcing the winner. The Grover-based protocol of Shi and Zhang \cite{shi2019privacy} marks bids by XOR, so equal bids cancel in pairs and a tied maximum is lost, its auctioneer learns intermediate bid values from Grover measurements, and its winner opens a commitment publicly. In the protocol of Hogg, Harsha, and Chen \cite{hogg2007quantum} the final measurement reveals the winner to the auctioneer, so it qualifies only if the auctioneer lies outside every coalition. The auctioneer-free protocol of Shi and Li \cite{shi2022feasible} qualifies after one modification, described below. The price phase of its classical analogue SEAL \cite{bag2020seal}, which computes the maximum bit by bit through anonymous veto, qualifies without modification under the decisional Diffie-Hellman assumption.

\emph{Price phase of Shi and Li.} Take bids in $\{0,\dots,B-1\}$ with $B=2^b$. At each of the $b$ levels, the current interval that contains the maximum is split into two halves, every bidder inputs the indicator that its bid lies in the upper half, and a quantum secure multiparty disjunction (SMD) protocol returns the OR of these inputs. A positive outcome selects the upper half and a negative outcome selects the lower half, and the final unit interval is $\{b_{\max}\}$. This is Steps 1 to 4 of the highest-bid phase of \cite{shi2022feasible} with $d=2$ intervals per level and recursion down to unit width. The second disjunction of each level is skipped because a negative first outcome already locates the maximum. We omit the commitment phase, Step 5, and the verification phase of \cite{shi2022feasible}. In Step 5 the winner claims the item publicly, which Proposition \ref{prop:pub} rules out, and the other two phases enforce fairness against malicious bidders, which lies outside the semi-honest model.

\emph{The SMD subroutine.} In each SMD round, $P_i$ prepares a Bell state $|\phi_{s_it_i}\rangle$ with private uniform features $s_it_i$ and passes one particle to $P_{i+1}$ around a ring. It applies the Pauli operator $U_{p_iq_i}$ with $p_i\vee q_i=x_i$ to one of the two particles it holds, measures those two particles in the Bell basis with outcome $\bar s_i\bar t_i$, and announces $y_i=s_it_i\oplus\bar s_i\bar t_i$. The disjunction returns 1 at the first round whose public parity $y=\bigoplus_iy_i$ differs from $00$, and returns 0 after $k$ rounds.

\begin{lemma}[Ring outcomes]\label{lem:ring}
For fixed features $s_it_i$ and Pauli labels $p_iq_i$, the $N$ Bell-measurement outcomes of an SMD round are uniformly distributed over the $4^{N-1}$ strings that satisfy $\bigoplus_i\bar s_i\bar t_i=\bigoplus_is_it_i\oplus\bigoplus_ip_iq_i$. In particular $y=\bigoplus_ip_iq_i$.
\end{lemma}

The parity relation is Theorem 3 of \cite{shi2022feasible}. Uniformity is proved in Appendix \ref{app:shili}, and we also confirmed it by exact state-vector computation for $N=4$ with random features, labels, and choices of the particle that carries each Pauli operator.

As specified in \cite{shi2022feasible}, a party with input 1 draws $p_iq_i$ uniformly from $\{01,10,11\}$. With $m$ unit inputs, $y$ is the XOR of $m$ independent uniform nonzero labels, and a character sum gives
\begin{equation}
\Pr[y=00]=\frac{1}{4}\Big[1+3\Big(-\frac{1}{3}\Big)^m\Big],
\label{eq:smdleak}
\end{equation}
which equals $0$, $1/3$, and $2/9$ for $m=1,2,3$ and coincides with the per-round error probability of \cite{shi2022feasible}. The public transcript of a disjunction therefore depends on $m$ and not only on the OR. For instance, with three parties and inputs $(0,1,1)$, a round with $y=00$ occurs with probability $1/3$ and tells $P_1$ that $x_2=x_3=1$. In the price phase this reveals how many bidders lie in a queried interval, a quantity that is not a function of the maximum. We therefore let a party with input 1 draw $p_iq_i$ uniformly from all four labels $\{00,01,10,11\}$. The XOR of $m\ge1$ independent labels that are uniform on all four values is uniform, so $\Pr[y=00]=1/4$ for every $m\ge1$, and a disjunction equal to 1 is missed after $k$ rounds with probability exactly $4^{-k}$.

\begin{proposition}[Shi and Li as the black box]\label{prop:shili}
Let $B=2^b\ge 32e\,\Shi$, let the parties be semi-honest, and use the price phase above with the modified labels and $k$ rounds per disjunction.
\begin{enumerate}[label=(\roman*),leftmargin=*]
\item One invocation returns the correct clearing price except with probability at most $\epsilon_k=b\,4^{-k}$, uniformly over the bids.
\item Without the $k$-round truncation, the price phase has max-bounded leakage.
\item Run $\Pi_{\rm race}$ with this black box, with the constants $14$ and $52$ replaced by $16$ and $93$, and with $k\ge\log_4(b\,M_{\max}/\epsilon)$ for some $\epsilon\le1$, where $M_{\max}=\lceil 16\ln(4(J+1)/\delta)\rceil+\lceil 93\,\Shi^2\ln(4/\delta)\rceil$. It outputs $S$ with probability at least $1-\delta$, and the view of every coalition is within total variation distance $\epsilon$ of the output of the simulator of Theorem \ref{thm:privacy}.
\item One invocation uses at most $bk$ SMD rounds with $N$ Bell states each, so the communication per invocation is $C_{\rm auc}(N)=O(kN\log B)$.
\end{enumerate}
\end{proposition}

The proof is given in Appendix \ref{app:shili}. Security against outside eavesdroppers relies on the authenticated quantum channels assumed in \cite{shi2022feasible}. We checked Proposition \ref{prop:shili} in a classical-level simulation of the price phase that samples the public parities from Lemma \ref{lem:ring}. With $B=64$ and $k=5$, the total variation distance between the joint laws of the clearing price and the number of $y=00$ rounds for $x=(8,0,0,0)$ and $x=(2,2,2,2)$ is $0.45$ with the published labels and $0.012$ with the modified labels, below the sampling floor of $0.03$ for $4\times10^5$ invocations. With $x=(3,0,4,1)$, $\Shi=8$, $B=256$, and $k=12$, the non-adaptive reduction recovered $S$ in $60$ out of $60$ runs.

\subsection{Randomness and optimality}\label{sec:optimality}

\begin{proposition}[Private randomness is necessary]\label{prop:det}
Let $N\ge2$ and $X\ge2$. Consider any protocol that runs rounds of $\Fp$ in which every bid is a deterministic function of the bidder's identity, its input, and the public price history, and whose price transcript depends on the inputs only through $S$. Then the transcript is identical for all inputs with $S\in\{1,\dots,X\}$. In particular, no such protocol computes $S$.
\end{proposition}

\begin{theorem}[Optimality for sampled bids]\label{thm:lower}
Consider any protocol that runs $M$ rounds of $\Fp$ and outputs a function of the public price transcript, in which, in every round, each party draws its bid independently from a distribution that may depend on its identity, its input, and the public price history, a party with input $0$ bids $0$, and the law of the price transcript depends on the inputs only through $S$. If the output equals the sum with probability at least $1-\delta$, where $\delta<1/4$, both on an input with sum $S\ge1$ and on an input with sum $S+1$, then
\begin{equation}
M\ge 2S^{2}\ln\frac{1}{4\delta}.
\label{eq:lower}
\end{equation}
\end{theorem}

Both proofs are given in Appendix \ref{app:lower}. The key step for Theorem \ref{thm:lower} is that privacy forces the conditional price law in every round to have the form $G(t)^S$ for a cumulative distribution $G$, which makes every admissible scheme a coarse-graining of the exponential race. The Kullback-Leibler divergence per round between sums $S$ and $S+1$ is then at most $1/S-\ln(1+1/S)\le 1/(2S^2)$. Theorems \ref{thm:correct} and \ref{thm:lower} together show that $\Theta(S^2\log(1/\delta))$ rounds are necessary and sufficient.

\begin{remark}[Scope of Theorem \ref{thm:lower}]
The convention that a zero input bids $0$ is essential. The complement encoding, in which $P_i$ uses rate $X+1-x_i$ in Eq.~\eqref{eq:key}, estimates $N(X+1)-S$ and needs $\Theta((N(X+1)-S)^2\log(1/\delta))$ rounds, which is fewer when $S>N(X+1)/2$. Whether any sampled scheme beats $\min\{S,\,N(X+1)-S\}^2$ is open. Theorem \ref{thm:lower} also excludes randomness correlated across parties and coherent access to the auction, which is treated next.
\end{remark}

\subsection{Coherent access to the auction}\label{sec:coherent}

Suppose now that the auction is available as a unitary together with its inverse, as in the distributed search architecture of Hogg, Harsha, and Chen \cite{hogg2007quantum}, where bid registers are passed between the auctioneer and the bidders without intermediate measurement. Let
\begin{equation}
A_{\rm price}:|b_1,\dots,b_N\rangle|0\rangle_P\mapsto|b_1,\dots,b_N\rangle\,|\max_i b_i\rangle_P ,
\end{equation}
and let party $P_i$ apply
\begin{equation}
V_i:|0\rangle\mapsto\sum_{b=0}^{B-1}\sqrt{\pi_{x_i}(b)}\,|b\rangle,\qquad \pi_x(b)=\Big(\frac{b+1}{B}\Big)^{x}-\Big(\frac{b}{B}\Big)^{x}\ (x\ge1),\quad \pi_0(b)=\mathbb{1}[b=0],
\end{equation}
which is the coherent version of the bid distribution of Eq.~\eqref{eq:key}. Write $\mathcal A=A_{\rm price}(V_1\otimes\cdots\otimes V_N)$ and let $\Pi_{<m}$ project the price register onto values below $m$. By Lemma \ref{lem:threshold}, $\|\Pi_{<m}\mathcal A|0\rangle\|^2=(m/B)^S$.

\begin{proposition}[Coherent reduction]\label{prop:coherent}
Let $S\le\tilde S\le\Shi$, for example from the scale stage of $\Pi_{\rm race}$, let $B\ge 8e\tilde S$, and let $m=\lfloor Be^{-1/\tilde S}\rfloor$. Amplitude estimation \cite{brassard2002quantum} of $\|\Pi_{<m}\mathcal A|0\rangle\|^2$ with $\Lambda=\lceil 36\tilde S\rceil$ returns an estimate $\tilde q$ with $\rnd(-\ln\tilde q/c)=S$ with probability at least $8/\pi^2$. Repeating $R$ times, with $R$ the smallest odd integer at least $5.2\ln(1/\delta)$, and returning the value produced by more than half of the runs yields $S$ with probability at least $1-\delta$, using at most $R(2\Lambda-1)=O(\tilde S\log(1/\delta))$ applications of $\mathcal A$ or $\mathcal A^\dagger$.
\end{proposition}
\begin{proof}
Amplitude estimation with $\Lambda$ evaluations returns $\tilde q$ with $|\tilde q-q|\le 2\pi\sqrt{q(1-q)}/\Lambda+\pi^2/\Lambda^2\le\pi/\Lambda+\pi^2/\Lambda^2\le 0.095/\tilde S$ with probability at least $8/\pi^2$ \cite{brassard2002quantum}. Since $0.095/\tilde S<\Delta$ by Eq.~\eqref{eq:delta} in Appendix \ref{app:thm1}, this accuracy guarantees correct rounding. The majority of $R$ independent runs is wrong with probability at most $\exp[-2R(8/\pi^2-1/2)^2]\le\delta$. Each run uses $\Lambda-1$ Grover iterations, each containing one $\mathcal A$ and one $\mathcal A^\dagger$, and one initial $\mathcal A$.
\end{proof}

Proposition \ref{prop:coherent} and Theorem \ref{thm:lower} separate sampled from coherent access to the same auction by a factor of order $S$ [Fig.~\ref{fig:reduction}(b)]. This separation is where the quantum setting matters. Every bidder must apply $V_i^\dagger$, the bid registers must never be measured before the final readout, and the encoded bids cannot be copied between Grover iterations. The privacy of the coherent version does not follow from Theorem \ref{thm:privacy}. The only measured outcome is the amplitude-estimation register, whose distribution depends on the inputs only through $(m/B)^S$, but the unmeasured registers pass through the auctioneer, so the appropriate notion is security against specious adversaries \cite{dupuis2010specious}. We leave a proof of privacy for the coherent version open.

\begin{figure*}[t]
\centering
\includegraphics[width=0.95\textwidth]{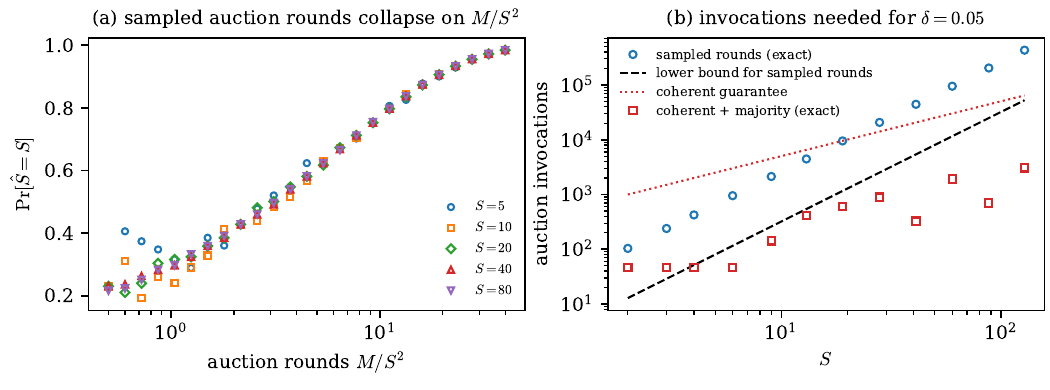}
\caption{Numerical verification of the reduction of Section \ref{auction_summation}, computed from exact binomial and exact amplitude-estimation output distributions. (a) Probability that the resolution stage returns the correct sum when $\tilde S=S$, as a function of the number of auction rounds in units of $S^2$, for $S=5$ to $80$. The curves collapse, confirming the $S^2$ scaling. (b) Auction invocations needed for failure probability $\delta=0.05$. Circles give the exact minimum number of sampled rounds, close to $26\,S^2$. The dashed line is the lower bound of Theorem \ref{thm:lower}, which applies to sampled rounds only. Squares give the exact number of applications of $\mathcal A$ or $\mathcal A^\dagger$ for amplitude estimation with majority voting, optimized over $\Lambda\in\{4S,\dots,64S\}$, and the dotted line is the guarantee of Proposition \ref{prop:coherent}. Monte Carlo simulations with explicit bidders reproduce Lemma \ref{lem:threshold} to within statistical error and give identical price distributions for $x=(3,0,5,2)$ and $x=(10,0,0,0)$.}
\label{fig:reduction}
\end{figure*}

\subsection{Leakage inherent to the sum}

By Theorem \ref{thm:privacy}, the reduction reveals nothing beyond $S$, so the residual leakage is that of $\Fsum$ itself. For independent inputs uniform on $\{0,\dots,X\}$ and $N\ge2$, the information that the sum carries about one input is
\begin{equation}
I(x_i;S)=H(S)-H(S\,|\,x_i)=H(S_N)-H(S_{N-1}),
\label{eq:sumleak}
\end{equation}
where $S_n$ denotes the sum of $n$ independent uniform inputs, because conditioning on $x_i$ shifts the remaining sum without changing its entropy. The leakage is strictly positive, since $S=0$ forces $x_i=0$ whereas $\Pr[x_i=0]=1/(X+1)$. Exact evaluation of Eq.~\eqref{eq:sumleak} with $X=99$ gives $0.721$, $0.316$, $0.163$, $0.076$, $0.037$, and $0.015$ bits for $N=2$, $3$, $5$, $10$, $20$, and $50$. For $N\ge5$ these values change by less than $3\%$ between $X=1$ and $X=999$, and they agree with $\frac12\log_2\frac{N}{N-1}\approx\frac{1}{2N\ln2}$ to within $2\%$. As $X\to\infty$ they converge to the difference of the differential entropies of the Irwin-Hall distributions of $N$ and $N-1$ uniform variables. The leakage therefore decays as $1/N$ and does not vanish as the input range grows. Reducing it requires changing the functionality, for example by adding noise to the sum, which is outside the scope of this work.

In the next section, we show that auction primitives can be constructed using only summation oracles.}

\section{Highest bid and winner from summation queries} \label{summation_auction}

{\rev
We now study the reverse direction. There are $N$ bidders with bids $b_1,\dots,b_N\in\{0,1,\dots,B_{\max}\}$, and the parties can run a secure summation protocol. Locating the highest bid through repeated summations of indicator functions is a selection from counting queries. The technique is standard, it is the structure of the quantum auction of Ref.~\cite{shi2021quantum}, and the disjunction-based auctions of Refs.~\cite{bag2020seal,shi2022feasible} follow the same pattern. We therefore do not present the search itself as a contribution. This section addresses four questions that the quantum setting raises: which query reveals the least, how the winner is found with the same primitive and without comparing bids with each other, what the consumption of quantum encodings by every query does to bidder incentives, and what an approximate quantum oracle costs inside an adaptive search. We also compare the construction with quantum maximum finding.

\subsection{Threshold disjunctions from summation}\label{sec:disj}

For a threshold $t$, the count $S_t=\sum_i\mathbb{1}[b_i\ge t]$ tells how many bidders reach $t$, but locating the maximum uses only whether $S_t>0$. Revealing the count therefore leaks information that the auction does not need (Section \ref{sec:countleak}). The disjunction alone is obtained from a summation protocol by randomizing the inputs.

\begin{lemma}[Disjunction from randomized summation]\label{lem:randsum}
Let $G$ be a finite abelian group and let a secure summation protocol output $\sum_iz_i\in G$. Suppose that a party with input $x_i=0$ submits $z_i=0$ and a party with input $x_i=1$ submits an independent uniform $z_i\in G$. Then the output equals $0$ with probability 1 if all inputs are 0, and it is uniform on $G$ otherwise. Declaring the disjunction to be $\mathbb{1}[\sum_iz_i\ne0]$ errs only when the disjunction is 1, with probability $1/|G|$. For every coalition $C$, its view, consisting of $z_C$ and the output, depends on the honest inputs only through $\bigvee_{i\in H}x_i$.
\end{lemma}
\begin{proof}
A sum that contains at least one independent uniform element of $G$ is uniform on $G$ and independent of the other summands. The coalition's view determines $\sum_{i\in H}z_i$, which is uniform if some honest input is 1 and equals $0$ otherwise.
\end{proof}

With $G=\mathbb{Z}_d$ the lemma uses summation modulo $d$, as computed by the protocol of Ref.~\cite{yang2018secure} with the corrections of Refs.~\cite{gu2019improvement,zhang2019improvements}. With $G=\mathbb{Z}_2\times\mathbb{Z}_2$ it is exactly the modified disjunction of Section \ref{sec:existing}, whose public parity is the sum of the Pauli labels.

\subsection{One search for the price and the winner}\label{sec:search}

Let $\beta=\lceil\log_2(B_{\max}+1)\rceil$, $\nu=\lceil\log_2N\rceil$, and $L=\beta+\nu$. Before the auction the parties draw a public uniformly random injection $\pi:\{1,\dots,N\}\to\{0,\dots,2^\nu-1\}$, and bidder $i$ forms the composite key $\kappa_i=b_i2^\nu+\pi(i)$. The keys are distinct, and their order is the order of the bids with ties broken by $\pi$. The search sets $P=0$ and, for $\ell=L-1,L-2,\dots,0$, queries the threshold $t=P+2^\ell$. Every bidder inputs $x_i=\mathbb{1}[\kappa_i\ge t]$ into the disjunction of Lemma \ref{lem:randsum}, and $P$ is replaced by $t$ if the disjunction is 1. The output $P$ gives the price $\lfloor P/2^\nu\rfloor$ and the winner $\pi^{-1}(P\bmod 2^\nu)$.

\begin{proposition}[Price and winner from one search]\label{prop:search}
\begin{enumerate}[label=(\roman*),leftmargin=*]
\item If no disjunction errs, the search returns $\kappa_{\max}=\max_i\kappa_i$. The price is $b_{\max}$, and the winner is the bidder with the largest $\pi$ among those bidding $b_{\max}$, which is uniformly distributed among them.
\item The search fails with probability at most $L/|G|$.
\item The public transcript is a function of $\kappa_{\max}$ and of the independent uniform sums returned by positive queries. The view of every coalition $C$ is a function of $\kappa_C$, $\max_{i\in H}\kappa_i$, and fresh randomness.
\item Each bidder compares only its own key with public thresholds. The search uses neither Grover search nor a test of $b_i=b_{\max}$, and it makes $L=O(\log(B_{\max}N))$ invocations of the summation protocol.
\end{enumerate}
\end{proposition}
\begin{proof}
(i) By induction on $\ell$, after the query at level $\ell$ the register $P$ equals $\kappa_{\max}$ with its lowest $\ell$ bits set to zero, because the disjunction at $t=P+2^\ell$ is 1 exactly when the $\ell$th bit of $\kappa_{\max}$ is 1. (ii) This follows from Lemma \ref{lem:randsum} and a union bound over the $L$ queries. (iii) The thresholds are functions of earlier outcomes. At threshold $t$ the honest disjunction equals $\mathbb{1}[\max_{i\in H}\kappa_i\ge t]$, and Lemma \ref{lem:randsum} gives the claim for each query. (iv) This holds by construction.
\end{proof}

The last $\nu$ levels of the search halve the set of bidders tied at $b_{\max}$ according to the bits of $\pi$. They are the subset-partitioning route, which stays inside the summation primitive. The statement in the Introduction that bids are not compared is meant in the precise sense of part (iv). Every bidder compares its own key with public thresholds, and no two bids are compared with each other.

\begin{remark}[Second-price payment]
After the search returns the winner $w$, a second search in which bidder $w$ inputs $0$ returns $\max_{j\ne w}\kappa_j$, whose price part is the second-highest bid, at a cost of $L$ further summations. The two transcripts reveal the highest bid as well, which an ideal second-price auction would hide. A refinement in the manner of the second-price extension of SEAL \cite{bag2020seal} hides the lower-order bits of the highest bid. The truthfulness of the second-price rule also presumes sealed bids, which the adaptive search does not provide (Proposition \ref{prop:adaptive}). Computing the second-highest bid alone requires the predicate $\mathbb{1}[S_t\ge2]$, which Lemma \ref{lem:randsum} does not supply.
\end{remark}

\subsection{Leakage of count-based searches}\label{sec:countleak}

If a search reveals the counts $S_t$ instead of the disjunctions, a linear scan over all thresholds reveals the multiplicities $n_k=S_k-S_{k+1}$ of every bid value, that is, the full bid histogram. For independent uniform bids every assignment of bids to bidders consistent with a histogram $T$ is equally likely, so the residual uncertainty is
\begin{equation}
H(\mathbf{b}\,|\,T)=\mathbb{E}_T\Big[\log_2\frac{N!}{\prod_kn_k!}\Big].
\end{equation}
A binary search that reveals counts discloses $S_t$ at every queried threshold. These counts depend on how many bidders lie above each threshold, which is not a function of $b_{\max}$. By Proposition \ref{prop:search}(iii) the disjunction search reveals $\kappa_{\max}$ and nothing else.

{\rev \emph{Information-theoretic form.} The contrast is an accounting of transcript entropy. Let a search use $q$ queries whose answers lie in an alphabet of size $m$, and let $T$ be the transcript, so that $H(T)\le q\log_2m$. Correctness requires $\kappa_{\max}$ to be a function of $T$. Writing $\mathbf{x}$ for the private inputs,
\begin{equation}
I(\mathbf{x};T)=I(\mathbf{x};\kappa_{\max})+I(\mathbf{x};T\,|\,\kappa_{\max})
\le I(\mathbf{x};\kappa_{\max})+q\log_2m-H(\kappa_{\max}),
\label{eq:budget}
\end{equation}
because $H(T\,|\,\kappa_{\max})=H(T)-H(\kappa_{\max})$. The quantity $q\log_2m-H(\kappa_{\max})$ is thus the budget for leakage beyond the outcome. The disjunction search spends none of it. It has $m=2$ and $q=L$, its transcript is the binary expansion of $\kappa_{\max}$, and the bound in Eq.~\eqref{eq:budget} is attained with the second term equal to zero. A linear scan that returns counts has $m=N+1$ and $q=B_{\max}$, so its budget is of order $B_{\max}\log_2N$ bits, and the budget is spent: the transcript fixes the histogram, whose residual uncertainty is given above. The relation between cost and privacy in this direction of the reduction is therefore not a trade-off. Queries with a smaller answer alphabet are both cheaper and less revealing, and the binary search wins on both counts. Trade-offs of the familiar kind appear in the other direction, where the number of auction rounds needed to resolve $S$ grows as the square of the accuracy demanded (Theorems \ref{thm:correct} and \ref{thm:lower}), and in the coherent version, where the accuracy of amplitude estimation improves only as the inverse of the number of invocations (Proposition \ref{prop:coherent}). The latter is the resolution bound of phase estimation, which is a statement about a single non-commuting pair, the phase and the number of applications of the Grover operator, rather than about two observables of the bidders.}

\subsection{Relation to quantum maximum finding}\label{sec:dh}

With coherent query access to the values $b_i$, D\"urr and H\o yer find an index of the maximum with $O(\sqrt N)$ queries \cite{durr1996quantum}. In the centralized query model this is better than the search of Proposition \ref{prop:search} with approximate quantum oracles, which needs a factor of order $L$ more queries (Section \ref{sec:budget}). The algorithm is not a substitute for a sealed-bid auction for three reasons. First, its oracle marks the indices $j$ with $b_j>b_y$ for the current candidate $y$, which compares the bids of two bidders and requires the candidate's bid to be available coherently to the others. Second, the candidates it measures form a chain of bidders with increasing bids, and this chain is revealed to whoever runs the algorithm although it is not a function of $(b_{\max},w)$. Third, in the circulating-register architecture used throughout this paper every coherent query requires every bidder to act, so $O(\sqrt N)$ queries cost $O(N^{3/2})$ bidder operations, compared with $O(NL)$ for Proposition \ref{prop:search} with an exact summation protocol. When a trusted party holds the bids coherently and leakage to that party is acceptable, quantum maximum finding is the better choice.

\subsection{Re-encoding and bidder strategy}\label{sec:strategy}

A classical answer can be stored and reused, but a quantum encoding of a bid is consumed by the query it serves and cannot be copied. Every query of Proposition \ref{prop:search} therefore takes fresh inputs, applied after the bidders have seen the outcomes of earlier queries. The search is a multi-round adaptive protocol, and this has strategic content.

\begin{proposition}[Adaptive bidding is undetectable]\label{prop:adaptive}
Let all bidders except bidder $i$ follow the search of Proposition \ref{prop:search} with their bids. Let bidder $i$ input $0$ at the $\beta$ price levels $\ell\ge\nu$, and at each index level input $\mathbb{1}[\hat b\,2^\nu+\pi(i)\ge t]$, where $\hat b=\lfloor P/2^\nu\rfloor$ is the price already fixed by the public outcomes. Then
\begin{enumerate}[label=(\roman*),leftmargin=*]
\item bidder $i$ learns $\max_{j\ne i}b_j$ before any of its inputs affects an outcome,
\item the public transcript has exactly the law it would have if bidder $i$ had honestly bid $\max_{j\ne i}b_j$, so no test on the public outcomes detects the deviation,
\item bidder $i$ wins at price $\max_{j\ne i}b_j$ whenever $\pi(i)$ exceeds $\pi(j)$ for every other bidder $j$ at that price, which has probability $1/(\tau+1)$ when $\tau$ other bidders are tied at the maximum.
\end{enumerate}
\end{proposition}
\begin{proof}
Let $m=\max_{j\ne i}b_j$ and let $\kappa'=m2^\nu+\pi(i)$ be the key of the honest bid $m$. A price-level threshold $t$ has its lowest $\nu$ bits equal to zero, so $\kappa'\ge t$ holds exactly when $m2^\nu\ge t$, which holds exactly when the key of the top other bidder reaches $t$. The disjunction is therefore unchanged by replacing the input $\mathbb{1}[\kappa'\ge t]$ with $0$, and by Lemma \ref{lem:randsum} so is the law of the returned sum. After the price levels $\hat b=m$, which gives (i), and the index-level inputs coincide with those of the honest key $\kappa'$, which gives (ii). Part (iii) follows from Proposition \ref{prop:search}(i) and the uniformity of $\pi$.
\end{proof}

A bidder thus conditions its effective bid on the other bids, which is exactly what a sealed-bid auction is designed to prevent. With four other bidders bidding independently and uniformly on $\{0,\dots,63\}$, a bidder with value $63$ obtains an expected utility of $5.03$ with its best fixed bid ($50$), and $6.06$ with the strategy of Proposition \ref{prop:adaptive}, in a simulation of $4\times10^5$ auctions. Inputs cannot be bound across queries with information-theoretic security by a commitment, because unconditionally secure quantum bit commitment is impossible \cite{mayers1997unconditionally,lo1997quantum}. Binding therefore requires computational assumptions, as in the zero-knowledge proofs of SEAL \cite{bag2020seal}, or relativistic constraints \cite{kent1999unconditionally,lunghi2013experimental}. Encoding all inputs in advance along a search tree does not help either, because inputs attached to tree nodes form a contingent strategy that is equivalent to adaptive play. A fixed query set does help.

\begin{proposition}[Sealed variant]\label{prop:sealed}
Suppose every bidder applies its inputs for all thresholds $t\in\{1,\dots,2^L-1\}$ before any outcome is revealed, and the outcomes are then revealed in decreasing order of $t$ until the first positive one, with $\kappa_{\max}=0$ if none is positive. This requires a summation protocol whose output step can be deferred for each instance, as in the disjunction of Section \ref{sec:existing}, where the announcement $y_i$ is the only input-dependent message of each party.
\begin{enumerate}[label=(\roman*),leftmargin=*]
\item With honest bidders the revealed outcomes determine $\kappa_{\max}$ and nothing else.
\item If bidder $i$ submits any inputs $x_i(t)$ while the others are honest, the first positive threshold is $\max(\max_{j\ne i}\kappa_j,K_i)$ with $K_i=\max\{t:x_i(t)=1\}$, which is the outcome of an honest bid with key $K_i$. A key $K_i$ whose last $\nu$ bits differ from $\pi(i)$ can only name another bidder at a price that bidder did not offer.
\item The protocol fails with probability at most $2^L/|G|$ and uses $2^L-1=O(B_{\max}N)$ summation instances in a single encoding round.
\end{enumerate}
\end{proposition}
\begin{proof}
With honest bidders the disjunction at $t$ is $\mathbb{1}[\kappa_{\max}\ge t]$, so the revealed outcomes are zeros above $\kappa_{\max}$ and a one at $\kappa_{\max}$. With bidder $i$ deviating, the disjunction at $t$ is $\mathbb{1}[\max_{j\ne i}\kappa_j\ge t]\vee x_i(t)$, whose largest positive threshold is $\max(\max_{j\ne i}\kappa_j,K_i)$. The inputs are fixed before any outcome is revealed, so they cannot depend on the others' bids. Part (iii) follows from Lemma \ref{lem:randsum} and a union bound.
\end{proof}

The descending reveal is the order of a Dutch auction, but the inputs are sealed, so bidder $i$ faces the strategic situation of a first-price sealed-bid auction \cite{krishna2002auction}. Sealed bidding without computational assumptions thus costs $O(B_{\max}N)$ summations in the constructions given here, against $O(\log(B_{\max}N))$ for the adaptive search. We do not know whether this gap is necessary.

\subsection{Error budget for approximate quantum oracles}\label{sec:budget}

A secure summation protocol over a group returns exact sums, and its only error in the search is the $1/|G|$ event of Lemma \ref{lem:randsum}. If a threshold query is instead answered by amplitude estimation of $a_t=S_t/N$ with the operator $U_f$ of the Notation section, the answer is approximate. We measure only whether the phase register of amplitude estimation \cite{brassard2002quantum} returns $0$.

\begin{proposition}[Adaptive search with amplitude estimation]\label{prop:budget}
Consider one threshold query answered by amplitude estimation with a $\Lambda$-dimensional phase register, declared positive when the register is nonzero.
\begin{enumerate}[label=(\roman*),leftmargin=*]
\item If $S_t=0$, the register returns $0$ with probability 1.
\item If $S_t\ge1$, the register returns $0$ with probability $\sin^2(\Lambda\theta)/(\Lambda^2\sin^2\theta)\le N/\Lambda^2$, where $\sin^2\theta=a_t$.
\item With $\Lambda=\lceil\sqrt{NL/\delta}\,\rceil$ at each of the $L$ queries of Proposition \ref{prop:search}, the search fails with probability at most $\delta$, and its transcript is within total variation distance $\delta$ of the transcript obtained with exact disjunctions. The search uses $O(L^{3/2}\sqrt{N/\delta})$ applications of the oracle, each requiring every bidder to act.
\item With $\Lambda=\lceil2\sqrt N\,\rceil$ and $R=\lceil\log_4(L/\delta)\rceil$ repetitions per query, declared positive if any repetition is nonzero, the failure probability is again at most $\delta$ with $O(L\sqrt N\log(L/\delta))$ oracle applications. The number of zero repetitions is then revealed, and its law depends on $S_t$.
\item If the ancilla is instead sampled directly, $s$ shots miss a positive query with probability at most $(1-1/N)^s$, so $s=\lceil N\ln(L/\delta)\rceil$ shots per query are needed.
\end{enumerate}
\end{proposition}
\begin{proof}
(i) If $a_t=0$, the prepared state lies in the eigenspace of the Grover operator with eigenvalue 1, so phase estimation returns $0$ with certainty. (ii) The state is an equal-weight superposition of eigenvectors with eigenphases $\pm2\theta$, and the amplitude of the outcome $0$ for each is $\Lambda^{-1}\sum_{j<\Lambda}e^{\pm2ij\theta}$, whose squared modulus is $\sin^2(\Lambda\theta)/(\Lambda^2\sin^2\theta)\le1/(\Lambda^2a_t)$. Since $a_t\ge1/N$, the bound follows. (iii) By (i) the only errors are misses of positive queries, each of probability at most $\delta/L$. A wrong answer can only derail the search, so the errors of the adaptive search do not compound beyond a union bound, and coupling with exact disjunctions gives the distance bound. (iv) Each repetition misses with probability at most $1/4$. (v) Each shot detects a positive query with probability $a_t\ge1/N$.
\end{proof}

Parts (iii) and (iv) scale as $\sqrt N$ while part (v) scales as $N$. This is the quadratic advantage of amplitude estimation over sampling, carried through an adaptive search with an explicit error budget. It is where interference enters the construction, and it is paid for in coherence: the bidders' unitaries must be applied $O(\sqrt N)$ times per query without intermediate measurement. We verified parts (i) and (ii) against the exact output distribution of amplitude estimation for $N$ up to $2048$, and Propositions \ref{prop:search} to \ref{prop:sealed} by simulating the search, the adaptive strategy, and all input vectors of a deviating bidder in a small sealed instance.
}

{\rev\section{Demonstrations on IBM Hardware} \label{sec:experiment}}

{\rev

\subsection{The search from summation queries}\label{sec:exp-search}

We executed the complete search of Proposition \ref{prop:search} on the 156-qubit Heron processor \texttt{ibm\_kingston} through Qiskit Runtime \cite{javadi2024qiskit}. The auction has $N=4$ bidders with bids in $\{0,1,2,3\}$, so $\beta=\nu=2$ and $L=4$. The public injection is $\pi=(2,0,3,1)$, and the keys $\kappa_i=4b_i+\pi(i)$ take values in $\{0,\dots,15\}$. Table \ref{tab:experiment} lists the five bid vectors. Three of them have ties at the top, and one has all bids equal to zero, so that the winner is decided by $\pi$ alone. The whole auction runs on one device, and each query circuit is therefore compiled from the inputs of all bidders. The experiment tests the query primitives and the search, not privacy.

\emph{Query primitives.} Figure \ref{fig:circuits} shows the two primitives for one threshold $t$.
\begin{enumerate}[label=(\alph*),leftmargin=*]
\item The disjunction query is Lemma \ref{lem:randsum} with $G=\mathbb{Z}_4$. Two qubits hold an element of $\mathbb{Z}_4$ and are prepared by the quantum Fourier transform. Every bidder with $\kappa_i\ge t$ adds an independent uniform $z_i\in\mathbb{Z}_4$ through the phase gates $P(2\pi z_i2^j/4)$ on qubit $j$, which translate the register by $z_i$ in the Fourier basis. The inverse transform then returns $\sum_iz_i \bmod 4$. Each threshold uses six rounds with fresh $z_i$ and $128$ shots per round. The value of a round is its most frequent outcome, and the query is declared positive if any round is nonzero, which misses a positive query with probability $4^{-6}$. After compilation the circuits have depth 29 to 34 and six two-qubit gates.
\item The sampling query is the direct readout of Proposition \ref{prop:budget}(v). Two index qubits are prepared in a uniform superposition over the bidders, and an oracle flips an ancilla for every bidder with $\kappa_i\ge t$, so that the ancilla reads $1$ with probability $S_t/N$. The oracle is compiled from the algebraic normal form of the indicator and needs at most one Toffoli gate and two CNOT gates. Each threshold uses $1024$ shots, and the query is declared positive if the frequency of $1$ exceeds $1/(2N)$. After compilation the circuits have depth 4 to 37 and at most eight two-qubit gates.
\item As a control, the disjunction query of bid vector I1 was repeated with a measurement of the register inserted after the bidders act. This removes the Fourier-basis phases that carry the inputs, so the output should become uniform and the search should fail.
\end{enumerate}

\emph{Execution.} All fifteen thresholds were executed in one batch for every bid vector, which is the fixed query set of Proposition \ref{prop:sealed}. The adaptive search of Proposition \ref{prop:search} reads the four queries on its path. Each query circuit depends only on $t$ and on fresh randomness, so this gives the same output distribution as issuing the four queries one after another. The descending reveal of Proposition \ref{prop:sealed} is read from the same data. The experiment used 615 circuits and $145\,920$ shots, without error mitigation.

\begin{figure*}[t]
\centering
\includegraphics[width=\textwidth]{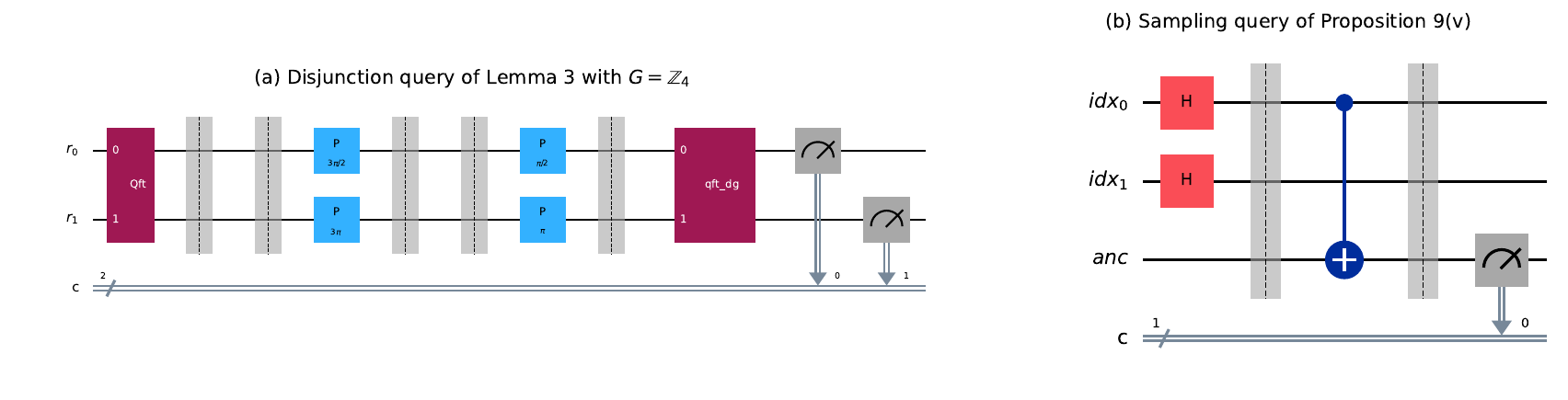}
\caption{Query circuits for bid vector I1 at threshold $t=12$, where bidders 2 and 4 have keys at or above the threshold. (a) Disjunction query of Lemma \ref{lem:randsum} with $G=\mathbb{Z}_4$. Each bidder acts between two barriers, here with $z=(0,3,0,1)$, and the register returns $\sum_iz_i\bmod4$. (b) Sampling query of Proposition \ref{prop:budget}(v). The ancilla reads $1$ with probability $S_t/N$, and for this threshold the oracle reduces to one CNOT gate.}
\label{fig:circuits}
\end{figure*}

\begin{table}[t]
\centering
\caption{Full searches on \texttt{ibm\_kingston}. For every bid vector, the adaptive search of Proposition \ref{prop:search} and the descending reveal of Proposition \ref{prop:sealed} returned the key $\kappa_{\max}$ listed, with both query primitives. The path gives the four thresholds of the adaptive search and their outcomes.}
\label{tab:experiment}
\begin{tabular}{lcccccc}
\hline\hline
 & Bids $b$ & Keys $\kappa$ & $\kappa_{\max}$ & Price & Winner & Adaptive path \\
\hline
I1 & $(2,3,1,3)$ & $(10,12,7,13)$ & 13 & 3 & 4 & $8^+\,12^+\,14^-\,13^+$ \\
I2 & $(1,0,2,0)$ & $(6,0,11,1)$ & 11 & 2 & 3 & $8^+\,12^-\,10^+\,11^+$ \\
I3 & $(0,0,0,0)$ & $(2,0,3,1)$ & 3 & 0 & 3 & $8^-\,4^-\,2^+\,3^+$ \\
I4 & $(3,3,3,3)$ & $(14,12,15,13)$ & 15 & 3 & 3 & $8^+\,12^+\,14^+\,15^+$ \\
I5 & $(1,2,0,2)$ & $(6,8,3,9)$ & 9 & 2 & 4 & $8^+\,12^-\,10^-\,9^+$ \\
\hline\hline
\end{tabular}
\end{table}

\begin{figure*}[t]
\centering
\includegraphics[width=\textwidth]{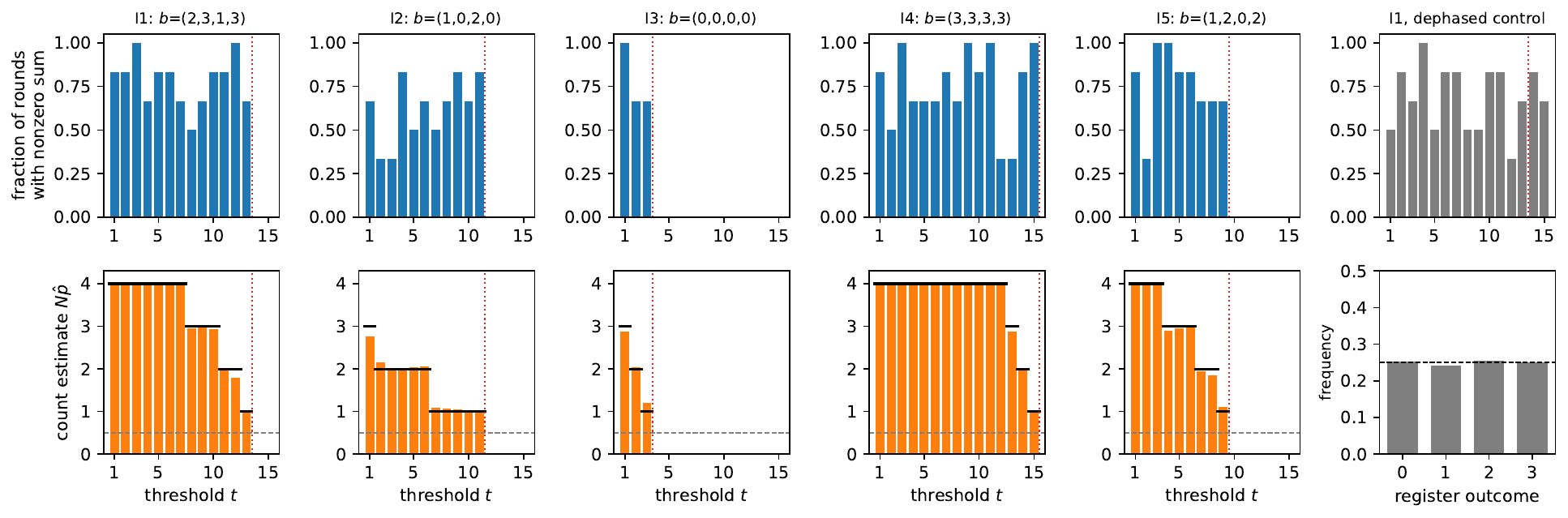}
\caption{Hardware results for all fifteen thresholds. Top row, columns 1 to 5: fraction of the six rounds of the disjunction query that returned a nonzero sum. Bottom row, columns 1 to 5: count estimate $N\hat p$ from the sampling query, with the true counts $S_t$ as black bars and the decision level $1/2$ as a dashed line. The dotted red line marks $\kappa_{\max}$. Column 6: the dephased control for I1, whose rounds are nonzero also above $\kappa_{\max}$ (top), and the pooled distribution of its register outcomes (bottom), which is uniform.}
\label{fig:hardware}
\end{figure*}

\emph{Results.} Both primitives returned the correct key $\kappa_{\max}$, and hence the correct price and winner, for all five bid vectors, through the adaptive search and through the descending reveal (Table \ref{tab:experiment}). None of the 75 disjunctions was wrong for either primitive. Figure \ref{fig:hardware} shows all queries.

For the disjunction query, the fraction of shots that deviated from the ideal sum was $4.7\%$ on average and at most $10.2\%$ in any round. The ideal sum was therefore the most frequent outcome in every one of the 450 rounds. No round of a negative query returned a nonzero value (0 of 144). The 306 rounds of positive queries returned the values 0, 1, 2, and 3 in 82, 80, 72, and 72 cases, consistent with the uniform law of Lemma \ref{lem:randsum} ($\chi^2=1.1$ with three degrees of freedom). This uniformity comes from the bidders' randomness, and the hardware preserved it. A positive query therefore reveals only that some key reaches $t$, and not how many do.

For the sampling query, the frequency of $1$ was at most $0.49\%$ for negative queries and at least $24.1\%$ for positive ones, far from the decision level $12.5\%$. Rounding $N\hat p$ recovered the count $S_t$ exactly in all 75 queries. The transcript of a count-revealing search thus determines the entire histogram of keys, as discussed in Section \ref{sec:countleak}, and here the whole bid vector up to the labels of the bidders.

In the control, the register outcomes were uniform on $\mathbb{Z}_4$. Their mean total variation distance from the uniform law was $0.062$ per round, which equals the value $0.061$ expected from sampling noise with 128 shots. The most frequent outcome differed from the ideal sum in 71 of 90 rounds, the queries at $t=14$ and $t=15$, where no key reaches the threshold, were declared positive, and the search returned $\kappa=15$ instead of $13$. The disjunction query therefore depends on the coherence of the register in the Fourier basis, whereas the sampling query does not.

\emph{Platform requirements.} The two primitives make different demands on hardware. The sampling query needs no coherence between the operations of different bidders. A device that routes one particle to a single bidder at a time and records a flag reproduces its statistics exactly, so a linear-optical sampler with path encoding is sufficient, and the requirements are low-loss routing and detector efficiency, which are strengths of photonic platforms. Its costs are the leakage of the count and the $O(N)$ shots per query. The disjunction query instead needs a register of dimension $d=|G|$ that keeps its relative phases while every bidder acts. The demonstration realizes that register as $\log_2d$ qubits on a gate-based processor. A photonic platform can carry it either the same way or as a single qudit of dimension $d$, in a time-bin or frequency-bin encoding, provided the relative phase is stable across all bidders. The control run shows what is lost when that phase information is destroyed. Carrying phase stability across all $N$ bidders, rather than the arithmetic of the summation, is therefore the obstacle that a photonic realization of the leakage-minimal search has to overcome.

\emph{Scope of the demonstration.} The experiment shows that the query primitives answer threshold disjunctions correctly on present hardware and that one search over composite keys returns the price and the winner, including ties. It does not test privacy, since one device holds all inputs, and it does not distribute the parties over a network. The four queries of the adaptive search were read from a batch rather than issued in sequence, which gives the same output distribution but does not demonstrate the re-encoding behaviour of Section \ref{sec:strategy}.

\subsection{The sum from repeated auctions}\label{sec:exp-reduction}

The other direction of the paper was executed on the same processor. Here the black box is the auctioneer-free auction of Shi and Li \cite{shi2022feasible} with the modification of Section \ref{sec:existing}, and the whole chain runs on hardware: the disjunction rounds of their protocol, the auctions built from them, and the reduction of Section \ref{auction_summation} built from the auctions.

\emph{Setup.} Two parties hold private inputs $x_i\in\{0,1,2\}$ and bids lie in $\{0,\dots,3\}$, so $B=4$. Party $i$ draws a private uniform $U_i$ and bids $b_i=\lfloor4U_i^{1/x_i}\rfloor$, and bids $0$ when $x_i=0$, so that $\Pr[b_{\max}<m]=(m/4)^S$ by Lemma \ref{lem:threshold}. The four input vectors $(0,0)$, $(1,0)$, $(1,1)$, and $(2,2)$ give $S=0$, $1$, $2$, and $4$.

\emph{Protocol.} One round of the disjunction subroutine of Ref.~\cite{shi2022feasible} uses one Bell pair per party. Party $i$ prepares $|\phi_{s_it_i}\rangle$, sends one particle to the next party around the ring, applies the Pauli $U_{p_iq_i}$ with $p_i\vee q_i$ equal to its input, measures the swapped pair in the Bell basis, and announces $y_i$. The public value $y=\bigoplus_iy_i$ equals $\bigoplus_ip_iq_i$. A level is positive at the first round with $y\ne00$, and at most $k=5$ rounds are used, so a positive level is missed with probability $4^{-k}$. An auction scans the bid space from the top, computing at level $m$ the disjunction of the indicators $\mathbb{1}[b_i=m]$, and the first positive level is the clearing price. The round circuits use $2N$ qubits and have depth $17$ with seven two-qubit gates after compilation. Each round was sampled with $16$ shots and its public value taken as the majority. In total, $2874$ rounds implemented $470$ auctions.

\emph{Results.} The individual Bell outcomes of a round are random and only their parity is fixed. The per-shot error of that parity was $12.4\%$ for two parties and $16.7\%$ for three, and the majority over $16$ shots was nevertheless correct in all $3354$ rounds. Every auction therefore returned the clearing price that its bids determine, with one exception among the $470$, in which all five rounds of a positive level returned $00$. That is the intrinsic error of the disjunction subroutine and not a hardware failure, and its rate $4^{-5}$ predicts $0.4$ such events in a run of this size. Figure \ref{fig:reduction-hw} shows the clearing-price distributions against the law of Lemma \ref{lem:threshold}. Maximum likelihood over the auctions gives $\hat S=0.00$ with interval $[0.00,0.06]$, $\hat S=0.90$ with $[0.75,1.08]$, $\hat S=2.19$ with $[1.84,2.58]$, and $\hat S=4.76$ with $[3.98,5.66]$, from $20$, $150$, $150$, and $150$ auctions, at the $95\%$ level by likelihood ratio. Every interval contains the true sum, so the auction, used as a black box, computed the sum on hardware.

\emph{The label rule of Section \ref{sec:existing}.} The last panel of Fig.~\ref{fig:reduction-hw} tests the modification directly. Three parties ran the round $120$ times with $m$ of them holding input $1$. With the labels as published, the public value never took the value $00$ for $m=1$ and took it in $31.7\%$ of the rounds for $m=2$, against the predicted $1/3$. This is the counterexample to Theorem 5 of Ref.~\cite{shi2022feasible} in measured form, since a party that sees $00$ learns that at least two of the others hold $1$. With the modified labels the law is uniform in both cases, and the total variation distance between $m=1$ and $m=2$ is $0.13$, the level expected from sampling noise at $120$ rounds per setting. The condition required by Theorem \ref{thm:privacy} is restored, as claimed.

\begin{figure*}[t]
\centering
\includegraphics[width=\textwidth]{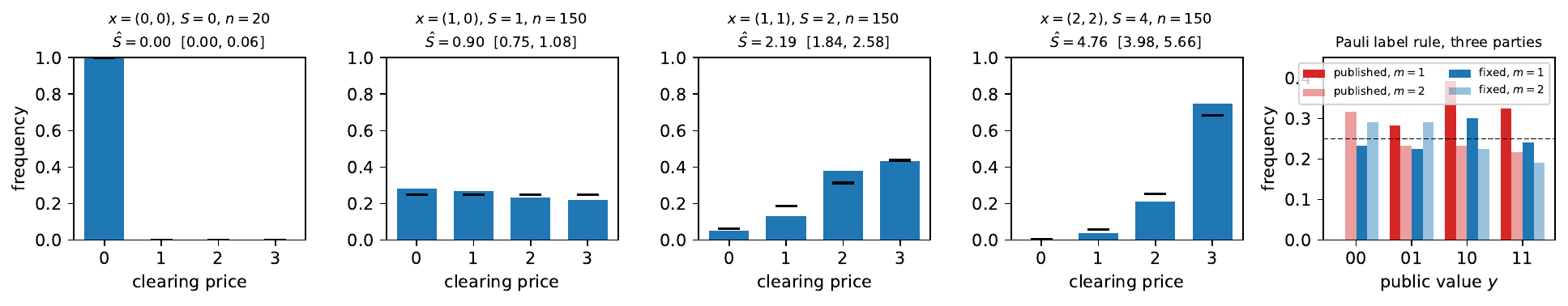}
\caption{The sum from repeated auctions on \texttt{ibm\_kingston}. First four panels: measured distribution of the clearing price for the four input vectors, with the law of Lemma \ref{lem:threshold} as black bars, and the maximum-likelihood estimate of $S$ with its $95\%$ interval. Last panel: law of the public value of one disjunction round with three parties, $m$ of which hold input $1$, under the labels of Ref.~\cite{shi2022feasible} and under the modification of Section \ref{sec:existing}. The published rule makes the law depend on $m$, and the modified rule does not.}
\label{fig:reduction-hw}
\end{figure*}

\emph{Scope.} As above, the parties are emulated on one device, so privacy is not tested and no network is involved. The parameters are small, $N=2$ and $B=4$ for the reduction, and the estimate of $S$ rests on the exact threshold law rather than on the asymptotic bounds of Theorem \ref{thm:correct}. Taken together, the two demonstrations execute one instance of each direction of the paper. The circuits, the raw counts, and the analysis scripts are available as stated below.
}

\section{Security Analysis}\label{sec:security-analysis}

{\rev
This section collects the security model and the security statements of the paper and states their limits.

\subsection*{Threat model}

Parties are semi-honest. They follow the protocol but may pool their views in a coalition of any size smaller than $N$, and they may use all information contained in the quantum states they hold and in the measurements they perform as prescribed. This model includes the honest-but-curious auctioneer considered in earlier quantum auction protocols and is stronger than a model without collusion. In Section \ref{sec:strategy} bidders may in addition choose their inputs strategically while following all other steps. Malicious parties that deviate from the prescribed operations, device-independent security, and side channels of imperfect hardware are outside the scope. Communication uses authenticated quantum and classical channels, and security against outside eavesdroppers is inherited from the protocols used as building blocks.

\subsection*{What is proved}

For summation from an auction, the view of every semi-honest coalition is exactly simulatable from its own inputs and the sum when the auction has max-bounded leakage (Theorem \ref{thm:privacy}). It is simulatable within total variation distance $\epsilon$ for the modified protocol of Shi and Li (Proposition \ref{prop:shili}) and within $M_{\max}S/B$ when only the winner is notified (Proposition \ref{prop:priv}). A public winner reveals every input (Proposition \ref{prop:pub}), and the coherent version of Proposition \ref{prop:coherent} has no privacy proof. For the auction from summation, the randomized inputs of Lemma \ref{lem:randsum} make the search reveal only the highest key, and coalition views depend only on the coalition's keys and the highest honest key (Proposition \ref{prop:search}). Searches that reveal counts leak more (Section \ref{sec:countleak}). With adaptive inputs, sealed bidding fails undetectably (Proposition \ref{prop:adaptive}), and a fixed query set restores it (Proposition \ref{prop:sealed}). With amplitude-estimation oracles, the transcript is within $\delta$ of the exact one (Proposition \ref{prop:budget}). The information about individual inputs that the sum itself carries is quantified in Section \ref{auction_summation}\,G.

\subsection*{What is not claimed}

Applying private operations to a shared register does not by itself ensure that only aggregate information is revealed. The sampling query of Section \ref{sec:experiment} is an example, because its statistics reveal the number of bidders above the threshold and not only the disjunction. Privacy therefore has to be established for each concrete protocol, as done above. The security of the quantum summation protocols used as building blocks is that established in their own analyses, including the attacks and corrections reported for Ref.~\cite{yang2018secure} in Refs.~\cite{gu2019improvement,zhang2019improvements}. Quantum sealed-bid auction protocols in the literature differ substantially in what they reveal (Section \ref{sec:existing}), so their security assumptions are not interchangeable. Hash commitments, such as the one in Ref.~\cite{shi2022feasible}, are at most computationally hiding and binding, and binding bids across adaptive queries with information-theoretic security is impossible (Section \ref{sec:strategy}).}

\section{Cost Analysis}\label{cost_analysis}

In this section, we analyze the computational, communication, and memory requirements associated with the proposed {\rev mutual reducibility} between quantum sealed-bid auction protocols and secure multi-party quantum summation protocols. We consider both directions of the reduction and compare the resulting complexities with those of representative existing protocols.

\subsection{Auction-to-Summation Reduction}

{\rev
The reduction of Section \ref{auction_summation} uses the auction strictly as a black box. Its cost is therefore the number of auction invocations multiplied by the cost $C_{\rm auc}(N)$ of one invocation of the chosen auction protocol, for instance $C_{\rm auc}(N)=O(kN\log B)$ for the instantiation of Proposition \ref{prop:shili}, where $k=O(\log(\Shi/\epsilon)+\log\log(1/\delta))$ suffices for leakage at most $\epsilon$. By Theorem \ref{thm:correct}, with $B=\Theta(\Shi)$, the total communication is
\begin{equation}
C_{A\rightarrow S}=O\!\Big(\big[\max(S,1)^2\log(1/\delta)+\log(\log\Shi/\delta)\big]\,C_{\rm auc}(N)\Big),
\end{equation}
which in the worst case $S=\Shi=NX$ is $O(N^2X^2\log(1/\delta)\,C_{\rm auc}(N))$. Each party holds one bid register per round, so that
\begin{equation}
Q_{A\rightarrow S}=\lceil\log_2 B\rceil=\log_2\Shi+O(1)
\end{equation}
qubits or bits per party. When the winner is notified privately (Proposition \ref{prop:priv}), the register grows by $O(\log\Shi+\log(1/\epsilon)+\log\log(1/\delta))$.

Theorem \ref{thm:lower} shows that the factor $S^2$ cannot be removed when bids are sampled independently. With coherent access to the auction (Proposition \ref{prop:coherent}) the number of applications of the auction unitary or its inverse drops to
\begin{equation}
T^{\rm coh}_{A\rightarrow S}=O\big(\max(S,1)\log(1/\delta)\big),
\end{equation}
plus the $O(\log(\log\Shi/\delta))$ sampled rounds of the scale stage. The price is that every bidder must apply $V_i^\dagger$ and that no intermediate measurement of the bid registers is allowed. This quadratic gap between sampled and coherent access is the auction counterpart of the precision advantage of amplitude estimation over classical sampling \cite{brassard2002quantum}.}

\subsection{Summation-to-Auction Reduction}

{\rev Table \ref{tab:costs} summarizes the cost of locating the price and the winner. Bidder operations count the number of times a bidder must act on a transmitted quantum system, which is the communication measure of the circulating-register architecture. For the search of Proposition \ref{prop:search} over $\mathbb{Z}_d$, the choice $d=\lceil L/\delta\rceil$ gives failure probability at most $\delta$, and each summation instance uses $O(N)$ bidder operations on $d$-dimensional systems.

\begin{table*}[t]
\caption{Cost and leakage of locating the price and the winner, with $L=\lceil\log_2(B_{\max}+1)\rceil+\lceil\log_2N\rceil$ and failure probability $\delta$. Invocations count summation instances, oracle applications, or shots.}
\label{tab:costs}
\begin{ruledtabular}
\small
\begin{tabular}{lcccc}
Method & Invocations & Bidder operations & Revealed & Bids bound\\
\hline
Adaptive, exact summation (Prop.~\ref{prop:search}) & $L$ & $O(NL)$ & $\kappa_{\max}$ & no (Prop.~\ref{prop:adaptive})\\
Sealed (Prop.~\ref{prop:sealed}) & $2^L-1$ & $O(B_{\max}N^2)$ & $\kappa_{\max}$ & yes\\
Adaptive, amplitude estimation (Prop.~\ref{prop:budget}) & $O(L^{3/2}\sqrt{N/\delta})$ & $O(L^{3/2}N^{3/2}\delta^{-1/2})$ & $\kappa_{\max}$ within $\delta$ & no\\
Adaptive, sampling (Prop.~\ref{prop:budget}) & $O(LN\log(L/\delta))$ & $O(LN^2\log(L/\delta))$ & $\kappa_{\max}$ within $\delta$ & no\\
Quantum maximum finding \cite{durr1996quantum} & $O(\sqrt N)$ & $O(N^{3/2})$ & chain of candidates & no\\
\end{tabular}
\end{ruledtabular}
\end{table*}

\emph{Where the overhead sits.} Exploring the bid space costs $\lceil\log_2(B_{\max}+1)\rceil$ summation instances, and determining the winner among the bidders tied at $b_{\max}$ costs the further $\lceil\log_2N\rceil$ levels of Proposition \ref{prop:search}, with $O(N)$ bidder operations per instance. Neither cost is avoidable by a search that only asks threshold questions, since the answer has $\log_2(B_{\max}+1)+\log_2N$ bits and each query returns one. In the other direction, Theorem \ref{thm:correct} needs $O(S^2\log(1/\delta))$ auction rounds and Theorem \ref{thm:lower} shows that this is optimal for independently sampled bids, while coherent access reduces it to $O(S\log(1/\delta))$ invocations by Proposition \ref{prop:coherent}. The two reductions are therefore not cost-preserving, and composing them multiplies the two overheads: a sum computed through auctions that are themselves built from summations costs $O(S^2\log(1/\delta))$ auctions, each of $O(N\log(B_{\max}N))$ summation instances. The mutual reducibility established here is a structural statement, which transfers protocols and security properties between the two tasks. It is not a statement that either task can be solved as cheaply through the other as it can be solved directly, and this is one reason the word equivalence is avoided.

The adaptive search with an exact summation protocol is the cheapest, and it matches the $O(N\log B)$ scaling reported for Ref.~\cite{shi2021quantum} up to the $\log N$ levels that locate the winner. Its drawback is Proposition \ref{prop:adaptive}, and removing that drawback without computational assumptions costs the sealed variant. Amplitude estimation improves on direct sampling by a factor of order $\sqrt N$ in bidder operations. Quantum maximum finding needs fewer queries than amplitude estimation, but it compares bids directly and reveals a chain of candidates.}

\subsection{Comparison with Existing Protocols}

The practical implications of the proposed {\rev mutual reducibility} depend strongly on the underlying auction and summation protocols used as building blocks. The quantum auction protocol of \cite{shi2021quantum} determines the winner through repeated evaluations of secure summation primitives over progressively refined bid intervals. For a bid space of size $B$, the complexity scales as $O(N\log B)$. Since summation already constitutes the fundamental primitive in that
construction, applying the present {\rev reduction} preserves the same
asymptotic behavior. In this case, the reduction provides a structural
reinterpretation rather than a computational improvement. {\rev The construction of Section \ref{summation_auction} shares this structure. What it adds relative to Ref.~\cite{shi2021quantum} is the leakage-minimal query of Lemma \ref{lem:randsum}, the location of the winner within the same search, the incentive analysis of Section \ref{sec:strategy}, and the error budget of Section \ref{sec:budget}.} By contrast, several existing quantum auction protocols incur quadratic
communication overheads. For example, the circular auction protocol of \cite{PathakAuction1} requires multiple exchanges of entangled states among communication sub-circles, resulting in communication complexity $ O(N^2)$. Similarly, auction schemes based on bidder-to-bidder confirmation procedures also exhibit quadratic scaling. Replacing such auction mechanisms with {\rev the search of Section \ref{sec:search} reduces the communication cost of locating the price and the winner to $O(N\log(B_{\max}N))$ (Table \ref{tab:costs}), an asymptotic improvement by a factor of order $N/\log(B_{\max}N)$.} The converse reduction does not necessarily provide a computational
advantage. Existing secure quantum summation protocols, including those
of \cite{yang2018secure}, and \cite{shi2016secure}, already achieve $
O(N) $ communication complexity. {\rev Implementing summation through an auction therefore multiplies the cost of one auction invocation by $O(\max(S,1)^2\log(1/\delta))$ sampled rounds, and Theorem \ref{thm:lower} shows that this overhead cannot be avoided when bids are sampled independently. The overhead is the price of using the auction strictly as a black box.}

{\rev The principal contribution of the mutual reductions is structural rather than a universal computational speedup, and it is confined to single-item first-price sealed-bid auctions with scalar bids. Within this class, the reductions allow security analyses, hardware implementations, and algorithmic improvements developed for one task to be transferred to the other, with the overheads quantified above. Any future improvement in the per-invocation cost of quantum auction protocols of this class translates into a corresponding improvement for auction-based secure summation, and vice versa. The reductions do not transfer to combinatorial or multi-unit auctions, or to payment rules and incentive properties beyond those treated in Section \ref{summation_auction}.}

\section{Conclusions}\label{sec:conclusion}

In this work, we have identified an inherent structural symmetry between the existing protocols for quantum sealed-bid auction and secure multi-party summation. Although these two classes of secure multi-party computation tasks have traditionally been studied independently and motivated by different application domains, our analysis reveals that they share a common operational structure. Specifically, we showed that {\rev for single-item sealed-bid auctions with scalar bids, highest bid identification and winner determination} can be systematically reduced to repeated applications of a summation oracle. {\rev Conversely, we showed that a first-price sealed-bid auction used as an unmodified black box computes a secure sum when each party bids a private random key whose law depends on its input. If the auction discloses only the clearing price, the transcript reveals nothing beyond the sum, even to colluding semi-honest parties, and the residual leakage is that of the sum itself, which necessarily leaks non-zero information about each individual input. If the auction publishes the identity of the winner, the same transcript reveals every input. The auctioneer-free quantum auction of Ref.~\cite{shi2022feasible} meets the required condition once its disjunction subroutine is modified so that it no longer reveals how many bidders lie in a queried interval.} {\rev In the other direction, one adaptive search over composite keys, whose queries are summations with randomized inputs, locates the highest bid and a winner and reveals nothing else, whereas searches that reveal counts disclose how many bidders lie above each queried threshold. Because quantum encodings are consumed by each query, bidders re-encode at every step, and a bidder can then condition its effective bid on the other bids without detection. A variant with a fixed query set restores sealed bidding at a cost of $O(B_{\max}N)$ summations, and binding the adaptive search instead requires computational or relativistic assumptions, since unconditionally secure quantum bit commitment is impossible.} We have also done the computational cost analysis of these reductions and compared it with the some of the representative existing protocols. {\rev Computing a sum through sampled auction rounds requires $\Theta(S^2\log(1/\delta))$ rounds, a bound that is tight for independently sampled bids, and private randomness is necessary. Coherent access to the auction lowers the cost to $O(S\log(1/\delta))$ invocations through amplitude estimation, which separates sampled from coherent use of the same auction.} {\rev An explicit error budget shows that approximate amplitude-estimation oracles add only a union bound over the queries and keep a quadratic advantage over direct sampling. Compared with quantum maximum finding, the search needs more queries in the centralized query model, but it never compares bids with each other, reveals no intermediate candidates, and needs $O(N\log(B_{\max}N))$ rather than $O(N^{3/2})$ bidder operations in a distributed setting.}

{\rev We also executed both directions on IBM superconducting hardware. The complete search recovered the highest bid and the winner for five bid vectors with both query primitives, and a dephased control showed that the disjunction query relies on coherence. Repeated auctions built from the auctioneer-free protocol of Ref.~\cite{shi2022feasible} returned the sum of the private inputs in every case, and the leakage of its published label rule, and its removal by our modification, were measured on the same device. The demonstrations run on one device and do not test privacy.} The framework developed here is protocol-agnostic and does not depend on a particular physical realization. We expect that this perspective will facilitate the design of modular, hardware-independent quantum SMC protocols and enable the transfer of techniques between {\rev single-item sealed-bid auctions} and secure aggregation. Future work may extend this approach to more general adversarial models and explore its applicability to other multi-party primitives beyond summation and auctions. {\rev Two questions left open by the present analysis are the privacy of the coherent reduction against specious adversaries and whether sampled schemes with correlated randomness can beat the $S^2$ round complexity. The scope of these results is single-item sealed-bid auctions with scalar bids and the first-price rule. Combinatorial auctions, in which valuations are set functions over bundles and winner determination is a set packing problem, general payment rules, and equilibrium properties lie outside the framework, and for bundles and correlated bids the protocol of Ref.~\cite{hogg2007quantum} remains the relevant quantum approach.}

\section*{Acknowledgment}
AP thanks the Department of Science and Technology, Government of India for the support provided through the National Quantum Mission (NQM). KS acknowledges the National Initiative on Undergraduate Science (NIUS), Homi Bhabha Centre for Science Education (HBCSE-TIFR), for providing the opportunity to participate in the NIUS program and for facilitating the research mentorship that contributed to this work.

\section*{Competing Interests}
Authors declare that they don't have any competing interests.
\section*{Data Availability}
{\rev The code used to produce the numerical results of Sections \ref{auction_summation} and \ref{summation_auction}, including Fig.~\ref{fig:reduction} and the simulations reported in Sections \ref{sec:existing} and \ref{sec:strategy}, and the circuits, raw counts, and IBM job identifiers of Section \ref{sec:experiment} are available from the corresponding author upon reasonable request.}
\section*{Authors' Contribution}
AP conceptualized the problem and supervised the work. KS performed all the analysis and computation. SM provided help in security analysis and quantifying the leakage of information. All the authors contributed equally in preparing the manuscript.

\appendix

{\rev
\section{Proofs for Section \ref{auction_summation}}\label{app:proofs}

\subsection{Proof of Theorem \ref{thm:correct}}\label{app:thm1}

\emph{Step 1: discretized thresholds.} Let $0<c^\ast\le1$, $B\ge 8e/c^\ast$, and $m=\lfloor Be^{-c^\ast}\rfloor$. Then $m\ge 7$ and $c=-\ln(m/B)\in[c^\ast,\,1.25\,c^\ast]$. Indeed, $m\le Be^{-c^\ast}$ gives $c\ge c^\ast$. From $m\ge Be^{-c^\ast}-1$ we get $c\le c^\ast-\ln(1-e^{c^\ast}/B)\le c^\ast+2e^{c^\ast}/B\le c^\ast+2e/B\le 1.25\,c^\ast$, using $-\ln(1-u)\le 2u$ for $0\le u\le 1/2$. Finally $m\ge B/e-1\ge 8/c^\ast-1\ge 7$. The scale stage uses $c^\ast_j=2^{-j}\ge 2^{-J}>1/(4\Shi)$, which is covered by $B\ge 32e\,\Shi$. The resolution stage uses $c^\ast=1/\tilde S\ge 1/\Shi$, which is covered by $B\ge 8e\,\Shi$.

\emph{Step 2: rounding tolerance.} Let $c^\ast=1/\tilde S$ with $S\le\tilde S$, $c\in[c^\ast,1.25c^\ast]$, and $q=e^{-cS}$. Define $\Delta=q(1-e^{-c/2})$. If $|\hat q-q|<\Delta$, then $\hat q>qe^{-c/2}=e^{-c(S+1/2)}$ and $\hat q<q(2-e^{-c/2})\le qe^{c/2}=e^{-c(S-1/2)}$, because $e^{u}+e^{-u}\ge2$. Hence $|-\ln\hat q/c-S|<1/2$ and the rounding in Eq.~\eqref{eq:estimator} returns $S$. Since $cS\le1.25$, $q\ge e^{-1.25}$. With $u=c/2\in[1/(2\tilde S),\,0.625/\tilde S]$ and $1-e^{-u}\ge u(1-u/2)\ge 0.6875\,u$ for $u\le0.625$,
\begin{equation}
\Delta\ \ge\ e^{-1.25}\times\frac{0.34375}{\tilde S}\ \ge\ \frac{0.0984}{\tilde S}.
\label{eq:delta}
\end{equation}

\emph{Step 3: resolution stage.} Conditional on $\tilde S\ge S$, the $M_2$ rounds are fresh. Hoeffding's inequality \cite{hoeffding1963} with $M_2\ge 52\tilde S^2\ln(4/\delta)$ gives $\Pr[|\hat q-q|\ge\Delta]\le 2\exp(-2M_2\Delta^2)\le 2\exp[-1.007\ln(4/\delta)]\le\delta/2$.

\emph{Step 4: scale stage.} For each $j$, $c_j\in[2^{-j},1.25\cdot2^{-j}]$ and $q_j=e^{-c_jS}$. If $S\le 2^{j-1}$ then $q_j\ge e^{-0.625}>0.535$. If $S\ge 2^{j+1}$ then $q_j\le e^{-2}<0.136$. The threshold $0.335$ lies more than $0.19$ away from both. With $M_1\ge 14\ln(4(J+1)/\delta)$, Hoeffding's inequality bounds the failure probability of each of the $J+1$ tests by $2\exp(-2M_1\cdot0.19^2)\le\delta/[2(J+1)]$. All tests use the same rounds, and a union bound gives total failure at most $\delta/2$. On the complementary event the following hold. (i) The test at $j=J$ passes, because $2^{J-1}\ge\Shi\ge S$, so $j^\ast$ is well defined. (ii) Passing at $j^\ast$ excludes $S\ge 2^{j^\ast+1}$, hence $S\le\tilde S$. (iii) If $j^\ast\ge1$, failing at $j^\ast-1$ excludes $S\le 2^{j^\ast-2}$, hence $\tilde S\le 2^{j^\ast+1}<8S$. If $j^\ast=0$, then $\tilde S\le2$. In all cases $\tilde S\le 8\max(S,1)$.

\emph{Step 5: conclusion.} The output is wrong only if the scale stage or the resolution stage fails, which has probability at most $\delta$. On the success event of the scale stage, $M_2\le 52\cdot64\max(S,1)^2\ln(4/\delta)+1$, which gives Eq.~\eqref{eq:rounds}. Since $\tilde S\le\Shi$ always, $M_2\le\lceil 52\Shi^2\ln(4/\delta)\rceil$. For the non-adaptive variant, $\tilde S=\Shi\ge S$ holds deterministically and Step 3 with $\ln(2/\delta)$ in place of $\ln(4/\delta)$ gives failure at most $\delta$. \hfill$\square$

\subsection{Proofs of Proposition \ref{prop:det} and Theorem \ref{thm:lower}}\label{app:lower}

\emph{Proof of Proposition \ref{prop:det}.} Suppose the price histories before round $r$ coincide for all inputs with $S\in\{1,\dots,X\}$, which holds trivially for $r=1$. Let $g_i$ be the bid map of $P_i$ in round $r$ for this history. Restrict attention to inputs supported on $P_1$ and $P_2$, and let $W=\max_{k\ge3}g_k(0)$ (with $W=-\infty$ if $N=2$), $u(a)=\max(g_1(a),W)$, $v(b)=\max(g_2(b),W)$. The price of the input $(a,b,0,\dots,0)$ is $\max(u(a),v(b))$, and privacy requires it to equal a function $\varphi(a+b)$. We show $\varphi(x)=\varphi(1)$ for $1\le x\le X$ by induction on $x$. For $x\ge2$, the decompositions $(x-1,1)$, $(1,x-1)$, and $(x,0)$ give $\varphi(x)=\max(u(x-1),v(1))$, $\varphi(x)\ge u(1)$, and $\varphi(x)\ge v(0)$. By the decompositions $(x-1,0)$ and $(0,1)$ of smaller sums, $u(x-1)\le\varphi(x-1)=\varphi(1)$ and $v(1)\le\varphi(1)$, so $\varphi(x)\le\varphi(1)$. Also $\varphi(1)=\max(u(1),v(0))\le\varphi(x)$. Every input with sum $S\le X$ has the same price as $(S,0,\dots,0)$ by privacy, namely $\varphi(1)$. The histories before round $r+1$ therefore coincide as well. \hfill$\square$

\emph{Proof of Theorem \ref{thm:lower}.} \emph{Structure.} Fix a round and a price history $h$ of positive probability. Let $F_{i,x}(t)=\Pr[b_i\le t\mid h]$ for $t\in\{0,\dots,B\}$ when $P_i$ has input $x$. By the zero-input convention $F_{i,0}\equiv1$, and by independence $\Pr[p\le t\mid h]=\prod_iF_{i,x_i}(t)$. Because the law of the transcript depends only on $S$, so does this conditional law, which we denote $\Phi_S$. The input with $x$ at $P_i$ and zeros elsewhere gives $F_{i,x}=\Phi_x$ for every $i$. The input with $a$ at $P_i$ and $b$ at $P_k$, $k\ne i$, gives $\Phi_a\Phi_b=\Phi_{a+b}$ for $a+b\le X$. With $G=\Phi_1$, induction yields $\Phi_x=G^x$ for $x\le X$, and for a general input $\prod_i G^{x_i}=G^S$. Here $G=F_{i,1}$ is a cumulative distribution function on $\{0,\dots,B\}$.

\emph{Coupling.} Let $V$ have cumulative distribution $v^S$ on $[0,1]$ and set $Y=\min\{t:V\le G(t)\}$, which is well defined because $G(B)=1$. As $G$ is nondecreasing, $Y\le t$ if and only if $V\le G(t)$, so $\Pr[Y\le t]=G(t)^S$. The conditional price is therefore the image of $V$ under a map that does not depend on $S$, and the same map applied to $V$ with law $v^{S+1}$ produces the conditional price for sum $S+1$. By the data-processing inequality, the per-round divergence satisfies
\begin{equation}
D\big(P_S(p\mid h)\,\big\|\,P_{S+1}(p\mid h)\big)\le D(V_S\|V_{S+1})=\frac1S-\ln\!\Big(1+\frac1S\Big)\le\frac{1}{2S^2},
\end{equation}
where the equality follows from the densities $Sv^{S-1}$ and $(S+1)v^{S}$ together with $\mathbb E[-\ln V_S]=1/S$, and the last step uses $\ln(1+u)\ge u-u^2/2$.

\emph{Chain rule and testing.} By the chain rule for the Kullback-Leibler divergence, the transcripts $T$ of $M$ rounds satisfy $D(P_S^T\|P_{S+1}^T)\le M/(2S^2)$. Let $A$ be the event that the output differs from $S$. Then $P_S(A)\le\delta$ and $P_{S+1}(A^c)\le\delta$, since on $A^c$ the output is $S\ne S+1$. The Bretagnolle-Huber inequality \cite{tsybakov2009} gives $2\delta\ge P_S(A)+P_{S+1}(A^c)\ge\frac12\exp[-M/(2S^2)]$, hence $M\ge 2S^2\ln(1/(4\delta))$. \hfill$\square$
\subsection{Proofs of Lemma \ref{lem:ring} and Proposition \ref{prop:shili}}\label{app:shili}

\emph{Proof of Lemma \ref{lem:ring}.} Up to a global phase, $|\phi_{ij}\rangle=(\mathbb{1}\otimes P_{ij})|\phi_{00}\rangle$ for a Pauli operator $P_{ij}$, and the labels compose by XOR up to phases, which gives Theorems 2 and 3 of \cite{shi2022feasible}. The $N$ prepared pairs and the $N$ measured pairs are two perfect matchings of the $2N$ particles whose union is a single cycle. Contracting a product of $|\phi_{00}\rangle$ states along such a cycle with a product of $\langle\phi_{00}|$ bras, with the Pauli operators inserted, gives $2^{-N}\,\mathrm{Tr}(Q_1Q_2\cdots Q_{2N})$, where the $Q$ are the inserted Pauli operators up to transposition. The trace equals $2$ times a phase when the product of the $Q$ is proportional to the identity, which is the stated parity relation, and vanishes otherwise. Every consistent outcome string therefore has probability $4^{-(N-1)}$, and there are $4^{N-1}$ of them. \hfill$\square$

\emph{Proof of Proposition \ref{prop:shili}.} (i) Each level uses one disjunction. It errs only if the OR of its inputs is 1 and all $k$ parities equal $00$, which with the modified labels has probability $4^{-k}$ for every configuration of inputs. A union bound over the $b$ levels gives $\epsilon_k$.

(ii) Fix one SMD round and a coalition $C$ with $H\ne\emptyset$. The coalition sees its own features, labels, and outcomes, together with the values $y_i$ announced by the honest parties. By Lemma \ref{lem:ring} the outcomes are uniform subject to one parity relation, and the honest features are uniform and private, so they act as one-time pads on the honest announcements. The coalition's round view is therefore uniform subject to the single relation $\bigoplus_{i\in C}\bar s_i\bar t_i\oplus\bigoplus_{i\in H}y_i=\bigoplus_{i\in C}s_it_i\oplus\bigoplus_ip_iq_i$. With the modified labels, $\bigoplus_{i\in H}p_iq_i$ is uniform when some honest input equals 1 and equals $00$ otherwise, so the round view depends on the honest inputs only through their OR, and the stopping round is a function of the round views. At a given level the current interval $[\ell,u)$ contains the global maximum, so no bid is $u$ or larger, and the OR of the honest inputs equals $\mathbb{1}[\max_{i\in H}b_i\ge(\ell+u)/2]$. The interval at each level is a function of earlier public outcomes. Composing the levels, the coalition's view of one invocation is a function of $b_C$, $\max_{i\in H}b_i$, and fresh randomness.

(iii) \emph{Correctness.} Since $\epsilon\le1$, part (i) gives $\epsilon_k\le1/M_{\max}$, and $M_{\max}\ge 93\ln4\,\Shi^2>50\,\Shi$, so $\epsilon_k<0.02/\Shi$. A wrong price affects only the indicator of its own invocation, so rounds stay independent and every threshold frequency has mean within $\epsilon_k$ of its value under Lemma \ref{lem:threshold}. In Step 3 of Appendix \ref{app:thm1} the Hoeffding margin becomes $\Delta-\epsilon_k\ge0.75\,\Delta$ by Eq.~\eqref{eq:delta}, and $2\cdot93\cdot(0.75\times0.0984)^2>1$. In Step 4 the margin becomes at least $0.1996-0.02>0.179$, and $2\cdot16\cdot0.179^2>1$. The failure bounds of Appendix \ref{app:thm1} therefore hold unchanged. \emph{Privacy.} Couple the protocol with a run that uses the same randomness but never truncates a disjunction. The two runs coincide unless some disjunction would need more than $k$ rounds, which by part (i) and a union bound over at most $M_{\max}$ invocations has probability at most $b\,M_{\max}4^{-k}\le\epsilon$. The untruncated run returns correct prices and has max-bounded leakage by part (ii), so Theorem \ref{thm:privacy} applies to it exactly.

(iv) There are $b$ levels and at most $k$ rounds per disjunction, and each round uses one Bell state and one Bell measurement per party. \hfill$\square$
}

\bibliographystyle{ieeetr}
\bibliography{main}

@article{preskill2018quantum, author={J. Preskill}, title={Quantum computing in the {NISQ} era and beyond}, journal={Quantum}, volume={2}, pages={79}, year={2018}, doi={10.22331/q-2018-08-06-79}}

@article{pirandola2020advances, author={S. Pirandola and U. L. Andersen and L. Banchi and M. Berta and D. Bunandar and R. Colbeck and D. Englund and T. Gehring and C. Lupo and C. Ottaviani and others}, title={Advances in quantum cryptography}, journal={Advances in Optics and Photonics}, volume={12}, number={4}, pages={1012--1236}, year={2020}, doi={10.1364/AOP.361502}}

@article{Shor, author={P. W. Shor}, title={Polynomial-time algorithms for prime factorization and discrete logarithms on a quantum computer}, journal={SIAM Journal on Computing}, volume={26}, pages={1484--1509}, year={1997}, doi={10.1137/S0097539795293172}}

@inproceedings{Grover, author={L. K. Grover}, title={A fast quantum mechanical algorithm for database search}, booktitle={Proceedings of the Twenty-Eighth Annual ACM Symposium on Theory of Computing}, pages={212--219}, publisher={ACM}, year={1996}, doi={10.1145/237814.237866}}

@book{QAlgo, author={A. M. Dalzell and S. McArdle and M. Berta and P. Bienias and C.-F. Chen and A. Gily{\'e}n and C. T. Hann and M. J. Kastoryano and E. T. Khabiboulline and A. Kubica and G. Salton and S. Wang and F. G. S. L. Brand{\~a}o}, title={Quantum Algorithms: A Survey of Applications and End-to-end Complexities}, publisher={Cambridge University Press}, year={2025}, doi={10.1017/9781009639651}}

@article{akpthk, author={A. Shenoy-Hejamadi and A. Pathak and S. Radhakrishna}, title={Quantum cryptography: Key distribution and beyond}, journal={Quanta}, volume={6}, number={1}, pages={1--47}, year={2017}, doi={10.12743/quanta.v6i1.57}}

@article{portmann2022security, author={C. Portmann and R. Renner}, title={Security in quantum cryptography}, journal={Reviews of Modern Physics}, volume={94}, number={2}, pages={025008}, year={2022}, doi={10.1103/RevModPhys.94.025008}}

@inproceedings{Crepeau2002, author={C. Cr{\'e}peau and D. Gottesman and A. Smith}, title={Secure multi-party quantum computation}, booktitle={Proceedings of the thiry-fourth annual ACM symposium on Theory of computing}, pages={643--652}, year={2002}, doi={10.1145/509907.510000}}

@inproceedings{GMW1987, author={O. Goldreich and S. Micali and A. Wigderson}, title={How to play any mental game}, booktitle={Proceedings of the Nineteenth Annual ACM Symposium on Theory of Computing}, pages={218--229}, publisher={ACM}, year={1987}, doi={10.1145/28395.28420}}

@inproceedings{Yao1982, author={A. C. Yao}, title={Protocols for secure computations}, booktitle={23rd annual symposium on foundations of computer science (sfcs 1982)}, pages={160--164}, publisher={IEEE}, year={1982}, doi={10.1109/SFCS.1982.38}}

@article{Lindell2017, author={Y. Lindell}, title={Secure multiparty computation}, journal={Communications of the ACM}, volume={64}, number={1}, pages={86--96}, year={2021}, doi={10.1145/3387108}}

@article{shi2021quantum, author={R.-H. Shi}, title={Quantum sealed-bid auction without a trusted third party}, journal={IEEE Transactions on Circuits and Systems I: Regular Papers}, volume={68}, number={10}, pages={4221--4231}, year={2021}, url={https://ieeexplore.ieee.org/document/9513292}}

@article{PathakAuction1, author={R. D. Sharma and K. Thapliyal and A. Pathak}, title={Quantum sealed-bid auction using a modified scheme for multiparty circular quantum key agreement}, journal={Quantum Information Processing}, volume={16}, number={7}, pages={169}, year={2017}, doi={10.1007/s11128-017-1620-0}}

@article{PathakAuction2, author={P. Asagodu and K. Thapliyal and A. Pathak}, title={Quantum and semi-quantum sealed-bid auction: vulnerabilities and advantages}, journal={Quantum Information Processing}, volume={21}, number={5}, pages={185}, year={2022}, doi={10.1007/s11128-022-03537-2}}

@article{Vaccaro2007Voting, author={J. A. Vaccaro and J. Spring and A. Chefles}, title={Quantum protocols for anonymous voting and surveying}, journal={Physical Review A}, volume={75}, pages={012333}, year={2007}, doi={10.1103/PhysRevA.75.012333}}

@article{PathakVoting1, author={K. Thapliyal and R. D. Sharma and A. Pathak}, title={Protocols for quantum binary voting}, journal={International Journal of Quantum Information}, volume={15}, number={01}, pages={1750007}, year={2017}, doi={10.1142/S0219749917500071}}

@article{PathakVeto2, author={S. Kumar and A. Pathak}, title={Experimental realization of quantum anonymous veto protocols using {IBM} quantum computer}, journal={Quantum Information Processing}, volume={21}, number={9}, pages={311}, year={2022}, doi={10.1007/s11128-022-03650-2}}

@misc{rahaman2015quantum, author={R. Rahaman and G. Kar}, title={{GHZ} correlation provides secure anonymous veto protocol}, howpublished={arXiv:1507.00592}, year={2015}, url={https://arxiv.org/abs/1507.00592}}

@article{PathakVeto1, author={S. Mishra and K. Thapliyal and A. Parakh and A. Pathak}, title={Quantum anonymous veto: a set of new protocols}, journal={EPJ Quantum Technology}, volume={9}, number={1}, pages={14}, year={2022}, doi={10.1140/epjqt/s40507-022-00133-2}}

@article{ji2019quantum, author={Z. Ji and H. Zhang and H. Wang and F. Wu and J. Jia and W. Wu}, title={Quantum protocols for secure multi-party summation}, journal={Quantum Information Processing}, volume={18}, number={6}, pages={168}, year={2019}, doi={10.1007/s11128-018-2141-1}}

@article{yang2018secure, author={H.-Y. Yang and T.-Y. Ye}, title={Secure multi-party quantum summation based on quantum fourier transform}, journal={Quantum Information Processing}, volume={17}, number={6}, pages={129}, year={2018}, doi={10.1007/s11128-018-1890-1}}

@article{shi2016secure, author={R.-h. Shi and Y. Mu and H. Zhong and J. Cui and S. Zhang}, title={Secure multiparty quantum computation for summation and multiplication}, journal={Scientific Reports}, volume={6}, number={1}, pages={19655}, year={2016}, doi={10.1038/srep19655}}

@article{PathakEcommerce, author={K. Thapliyal and A. Pathak}, title={Quantum e-commerce: a comparative study of possible protocols for online shopping and other tasks related to e-commerce}, journal={Quantum Information Processing}, volume={18}, number={6}, pages={191}, year={2019}, doi={10.1007/s11128-019-2309-3}}

@article{mishra2023quantum, author={S. Mishra and A. Pathak}, title={Quantum and semi-quantum lottery: strategies and advantages}, journal={Quantum Information Processing}, volume={22}, number={7}, pages={290}, year={2023}, doi={10.1007/s11128-023-04041-x}}

@misc{Goldreich2009SMC, author={O. Goldreich}, title={Secure multi-party computation}, howpublished={Unpublished manuscript, final revision October 2002}, year={2002}, url={https://www.wisdom.weizmann.ac.il/~oded/pp.html}}

@article{brassard2002quantum, author={G. Brassard and P. H{\o}yer and M. Mosca and A. Tapp}, title={Quantum amplitude amplification and estimation}, journal={Contemporary Mathematics}, volume={305}, pages={53--74}, year={2002}, doi={10.1090/conm/305/05215}}

@article{grinko2021iterative, author={D. Grinko and J. Gacon and C. Zoufal and S. Woerner}, title={Iterative quantum amplitude estimation}, journal={npj Quantum Information}, volume={7}, number={1}, pages={52}, year={2021}, doi={10.1038/s41534-021-00379-1}}

@article{dou2024quantum, author={Z. Dou and Y. Wang and Z. Liu and J. Bi and X. Chen and L. Li}, title={Quantum secure multi-party computational geometry based on multi-party summation and multiplication}, journal={Quantum Science and Technology}, volume={9}, number={2}, pages={025023}, year={2024}, doi={10.1088/2058-9565/ad34f5}}

@article{li2024secure, author={X. Li and Y. Xiong and C. Zhang}, title={Secure multiparty quantum computation for summation and data sorting}, journal={Quantum Information Processing}, volume={23}, number={9}, pages={321}, year={2024}, doi={10.1007/s11128-024-04528-1}}

@article{zhang2024new, author={Y. Zhang and Y. Yao and H. Sun and K. Zhang and T. Song}, title={A new hybrid protocol that simultaneously achieves quantum multiparty summation and ranking}, journal={Advanced Quantum Technologies}, volume={7}, number={6}, pages={2400078}, year={2024}, doi={10.1002/qute.202400078}}

@article{zhang2025novel, author={K. Zhang and Y. Zhang and X. Zhang and H. Liu and T. Song and G. Du}, title={A novel quantum multiparty summation protocol based on a cooperative random number mechanism}, journal={EPJ Quantum Technology}, volume={12}, number={1}, pages={57}, year={2025}, doi={10.1140/epjqt/s40507-025-00361-2}}

@article{guha2008quantum, author={S. Guha and T. Hogg and D. Fattal and T. Spiller and R. G. Beausoleil}, title={Quantum auctions using adiabatic evolution: The corrupt auctioneer and circuit implementations}, journal={International Journal of Quantum Information}, volume={6}, number={04}, pages={815--839}, year={2008}, doi={10.1142/S0219749908004183}}

@article{hogg2007quantum, author={T. Hogg and P. Harsha and K.-Y. Chen}, title={Quantum auctions}, journal={International Journal of Quantum Information}, volume={5}, number={5}, pages={751--780}, year={2007}, doi={10.1142/S0219749907003183}}

@article{cohen1997size, author={E. Cohen}, title={Size-estimation framework with applications to transitive closure and reachability}, journal={Journal of Computer and System Sciences}, volume={55}, number={3}, pages={441--453}, year={1997}, doi={10.1006/jcss.1997.1534}}

@article{mosk2008fast, author={D. Mosk-Aoyama and D. Shah}, title={Fast distributed algorithms for computing separable functions}, journal={IEEE Transactions on Information Theory}, volume={54}, number={7}, pages={2997--3007}, year={2008}, doi={10.1109/TIT.2008.924648}}

@article{efraimidis2006weighted, author={P. S. Efraimidis and P. G. Spirakis}, title={Weighted random sampling with a reservoir}, journal={Information Processing Letters}, volume={97}, number={5}, pages={181--185}, year={2006}, doi={10.1016/j.ipl.2005.11.003}}

@article{hoeffding1963, author={W. Hoeffding}, title={Probability inequalities for sums of bounded random variables}, journal={Journal of the American Statistical Association}, volume={58}, number={301}, pages={13--30}, year={1963}, doi={10.1080/01621459.1963.10500830}}

@book{tsybakov2009, author={A. B. Tsybakov}, title={Introduction to Nonparametric Estimation}, publisher={Springer}, address={New York}, year={2009}, doi={10.1007/b13794}}

@inproceedings{dupuis2010specious, author={F. Dupuis and J. B. Nielsen and L. Salvail}, title={Secure two-party quantum evaluation of unitaries against specious adversaries}, booktitle={Advances in Cryptology, CRYPTO 2010}, series={Lecture Notes in Computer Science}, volume={6223}, pages={685--706}, publisher={Springer}, year={2010}, doi={10.1007/978-3-642-14623-7_37}}

@article{shi2022feasible, author={R.-H. Shi and Y.-F. Li}, title={A feasible quantum sealed-bid auction scheme without an auctioneer}, journal={IEEE Transactions on Quantum Engineering}, volume={3}, pages={2100212}, year={2022}, doi={10.1109/TQE.2022.3148664}}

@article{bag2020seal, author={S. Bag and F. Hao and S. F. Shahandashti and I. G. Ray}, title={{SEAL}: Sealed-bid auction without auctioneers}, journal={IEEE Transactions on Information Forensics and Security}, volume={15}, pages={2042--2052}, year={2020}, doi={10.1109/TIFS.2019.2955793}}

@article{shi2019privacy, author={R.-H. Shi and M. Zhang}, title={Privacy-preserving quantum sealed-bid auction based on {G}rover's search algorithm}, journal={Scientific Reports}, volume={9}, pages={7626}, year={2019}, doi={10.1038/s41598-019-44030-8}}

@misc{durr1996quantum, author={C. D{\"u}rr and P. H{\o}yer}, title={A quantum algorithm for finding the minimum}, howpublished={arXiv:quant-ph/9607014}, year={1996}, url={https://arxiv.org/abs/quant-ph/9607014}}

@article{mayers1997unconditionally, author={D. Mayers}, title={Unconditionally secure quantum bit commitment is impossible}, journal={Physical Review Letters}, volume={78}, pages={3414}, year={1997}, doi={10.1103/PhysRevLett.78.3414}}

@article{lo1997quantum, author={H.-K. Lo and H. F. Chau}, title={Is quantum bit commitment really possible?}, journal={Physical Review Letters}, volume={78}, pages={3410}, year={1997}, doi={10.1103/PhysRevLett.78.3410}}

@article{kent1999unconditionally, author={A. Kent}, title={Unconditionally secure bit commitment}, journal={Physical Review Letters}, volume={83}, pages={1447}, year={1999}, doi={10.1103/PhysRevLett.83.1447}}

@article{lunghi2013experimental, author={T. Lunghi and J. Kaniewski and F. Bussi{\`e}res and R. Houlmann and M. Tomamichel and A. Kent and N. Gisin and S. Wehner and H. Zbinden}, title={Experimental bit commitment based on quantum communication and special relativity}, journal={Physical Review Letters}, volume={111}, pages={180504}, year={2013}, doi={10.1103/PhysRevLett.111.180504}}

@book{krishna2002auction, author={V. Krishna}, title={Auction Theory}, edition={2}, publisher={Academic Press}, address={San Diego}, year={2009}, url={https://www.sciencedirect.com/book/9780123745071/auction-theory}}

@article{chen2008experiments, author={K.-Y. Chen and T. Hogg}, title={Experiments with probabilistic quantum auctions}, journal={Quantum Information Processing}, volume={7}, pages={139--152}, year={2008}, doi={10.1007/s11128-008-0079-4}}

@article{piotrowski2003quantum, author={E. W. Piotrowski and J. S{\l}adkowski}, title={Quantum {E}nglish auctions}, journal={Physica A}, volume={318}, number={3--4}, pages={505--515}, year={2003}, doi={10.1016/S0378-4371(02)01533-9}}

@article{piotrowski2008quantum, author={E. W. Piotrowski and J. S{\l}adkowski}, title={Quantum auctions: Facts and myths}, journal={Physica A}, volume={387}, number={15}, pages={3949--3953}, year={2008}, doi={10.1016/j.physa.2008.02.071}}

@article{rothkopf1998computationally, author={M. H. Rothkopf and A. Pekec and R. M. Harstad}, title={Computationally manageable combinational auctions}, journal={Management Science}, volume={44}, number={8}, pages={1131--1147}, year={1998}, doi={10.1287/mnsc.44.8.1131}}

@article{sandholm2002algorithm, author={T. Sandholm}, title={Algorithm for optimal winner determination in combinatorial auctions}, journal={Artificial Intelligence}, volume={135}, number={1--2}, pages={1--54}, year={2002}, doi={10.1016/S0004-3702(01)00159-X}}

@book{milgrom2004putting, author={P. Milgrom}, title={Putting Auction Theory to Work}, publisher={Cambridge University Press}, address={Cambridge}, year={2004}, doi={10.1017/CBO9780511813825}}

@book{cramton2006combinatorial, editor={P. Cramton and Y. Shoham and R. Steinberg}, title={Combinatorial Auctions}, publisher={MIT Press}, address={Cambridge, MA}, year={2006}, doi={10.7551/mitpress/9780262033428.001.0001}}

@article{naseri2009secure, author={M. Naseri}, title={Secure quantum sealed-bid auction}, journal={Optics Communications}, volume={282}, number={9}, pages={1939--1943}, year={2009}, doi={10.1016/j.optcom.2009.01.026}}

@article{qin2009cryptanalysis, author={S.-J. Qin and F. Gao and Q.-Y. Wen and L.-M. Meng and F.-C. Zhu}, title={Cryptanalysis and improvement of a secure quantum sealed-bid auction}, journal={Optics Communications}, volume={282}, number={19}, pages={4014--4016}, year={2009}, url={https://www.sciencedirect.com/science/article/abs/pii/S003040180900649X}}

@article{yang2009improved, author={Y.-G. Yang and M. Naseri and Q.-Y. Wen}, title={Improved secure quantum sealed-bid auction}, journal={Optics Communications}, volume={282}, number={20}, pages={4167--4170}, year={2009}, doi={10.1016/j.optcom.2009.07.010}}

@article{zhao2010secure, author={Z. Zhao and M. Naseri and Y. Zheng}, title={Secure quantum sealed-bid auction with post-confirmation}, journal={Optics Communications}, volume={283}, number={16}, pages={3194--3197}, year={2010}, url={https://www.sciencedirect.com/science/article/abs/pii/S0030401810003433}}

@article{liu2016multiparty, author={W.-J. Liu and H.-B. Wang and G.-L. Yuan and Y. Xu and Z.-Y. Chen and X.-X. An and F.-G. Ji and G. T. Gnitou}, title={Multiparty quantum sealed-bid auction using single photons as message carrier}, journal={Quantum Information Processing}, volume={15}, number={2}, pages={869--879}, year={2016}, doi={10.1007/s11128-015-1202-y}}

@article{shi2022anonymous, author={R.-H. Shi}, title={Anonymous quantum sealed-bid auction}, journal={IEEE Transactions on Circuits and Systems II: Express Briefs}, volume={69}, number={2}, pages={414--418}, year={2022}, doi={10.1109/TCSII.2021.3098755}}

@article{ye2023measurement, author={C.-Q. Ye and J. Li and X.-B. Chen and Y. Hou and M. Dong and K. Ota}, title={Measurement-based quantum sealed-bid auction}, journal={IEEE Transactions on Circuits and Systems I: Regular Papers}, volume={70}, number={12}, pages={5352--5365}, year={2023}, url={https://ieeexplore.ieee.org/document/10272383}}

@article{mihara2023quantum, author={T. Mihara}, title={Quantum sealed-bid auction using the phases of quantum entangled states as bids}, journal={The European Physical Journal Plus}, volume={138}, pages={31}, year={2023}, doi={10.1140/epjp/s13360-023-03662-6}}

@article{franklin1996design, author={M. K. Franklin and M. K. Reiter}, title={The design and implementation of a secure auction service}, journal={IEEE Transactions on Software Engineering}, volume={22}, number={5}, pages={302--312}, year={1996}, doi={10.1109/32.502223}}

@inproceedings{naor1999privacy, author={M. Naor and B. Pinkas and R. Sumner}, title={Privacy preserving auctions and mechanism design}, booktitle={Proceedings of the 1st ACM Conference on Electronic Commerce}, pages={129--139}, year={1999}, doi={10.1145/336992.337028}}

@article{brandt2006obtain, author={F. Brandt}, title={How to obtain full privacy in auctions}, journal={International Journal of Information Security}, volume={5}, number={4}, pages={201--216}, year={2006}, doi={10.1007/s10207-006-0001-y}}

@inproceedings{bogetoft2009secure, author={P. Bogetoft and D. L. Christensen and I. Damg{\aa}rd and M. Geisler and T. Jakobsen and M. Kr{\o}igaard and J. D. Nielsen and J. B. Nielsen and K. Nielsen and J. Pagter and M. Schwartzbach and T. Toft}, title={Secure multiparty computation goes live}, booktitle={Financial Cryptography and Data Security}, series={Lecture Notes in Computer Science}, volume={5628}, pages={325--343}, publisher={Springer}, year={2009}, doi={10.1007/978-3-642-03549-4_20}}

@article{chen2010efficient, author={X.-B. Chen and G. Xu and Y.-X. Yang and Q.-Y. Wen}, title={An efficient protocol for the secure multi-party quantum summation}, journal={International Journal of Theoretical Physics}, volume={49}, number={11}, pages={2793--2804}, year={2010}, doi={10.1007/s10773-010-0472-5}}

@misc{gu2019improvement, author={J. Gu and T. Hwang}, title={Improvement on ``{S}ecure multi-party quantum summation based on quantum {F}ourier transform''}, howpublished={arXiv:1907.02656}, year={2019}, url={https://arxiv.org/abs/1907.02656}}

@article{zhang2019improvements, author={C. Zhang and M. Razavi and Z.-W. Sun and H.-Z. Situ}, title={Improvements on ``{S}ecure multi-party quantum summation based on quantum {F}ourier transform''}, journal={Quantum Information Processing}, volume={18}, number={11}, pages={336}, year={2019}, doi={10.1007/s11128-019-2449-5}}

@article{huang2014quantum, author={W. Huang and Q.-Y. Wen and B. Liu and Q. Su and S.-J. Qin and F. Gao}, title={Quantum anonymous ranking}, journal={Physical Review A}, volume={89}, pages={032325}, year={2014}, doi={10.1103/PhysRevA.89.032325}}

@article{heinrich2002quantum, author={S. Heinrich}, title={Quantum summation with an application to integration}, journal={Journal of Complexity}, volume={18}, number={1}, pages={1--50}, year={2002}, doi={10.1006/jcom.2001.0629}}

@misc{javadi2024qiskit, author={A. Javadi-Abhari and M. Treinish and K. Krsulich and C. J. Wood and J. Lishman and J. Gacon and S. Martiel and P. D. Nation and L. S. Bishop and A. W. Cross and B. R. Johnson and J. M. Gambetta}, title={Quantum computing with {Qiskit}}, howpublished={arXiv:2405.08810}, year={2024}, url={https://arxiv.org/abs/2405.08810}}

\end{document}